%% file: main.tex
\documentclass{article}

\usepackage{microtype}
\usepackage{graphicx}
\usepackage{subfigure}
\usepackage{subcaption}
\usepackage{booktabs} 
\usepackage{multirow}
\usepackage{arydshln}
\usepackage{booktabs}

\usepackage{hyperref}

\usepackage[accepted]{icml2024}

\usepackage{amsmath}
\usepackage{amssymb}
\usepackage{mathtools}
\usepackage{amsthm}

\usepackage[capitalize,noabbrev]{cleveref}

\theoremstyle{plain}

\theoremstyle{definition}

\theoremstyle{remark}

\usepackage[textsize=tiny]{todonotes}

\usepackage{soul}
\usepackage{paralist}

\usepackage{titlesec}
\titlespacing*{\section}{0pt}{0.75\baselineskip}{0.5\baselineskip}

\usepackage{inconsolata}

\titlespacing{\paragraph}{%
  0pt}{
  0.1\baselineskip}{
  1em}%

\icmltitlerunning{\textsc{Repoformer}: Selective Retrieval for Repository-Level Code Completion}

\begin{document}

\setlength{\abovedisplayskip}{5pt}
\setlength{\belowdisplayskip}{5pt}

\twocolumn[
\icmltitle{\textsc{Repoformer}: Selective Retrieval for Repository-Level Code Completion}



\icmlsetsymbol{equal}{*}

\begin{icmlauthorlist}
\icmlauthor{Di Wu}{sch,equal}
\icmlauthor{Wasi Uddin Ahmad}{comp}
\icmlauthor{Dejiao Zhang}{comp}
\icmlauthor{Murali Krishna Ramanathan}{comp}
\icmlauthor{Xiaofei Ma}{comp}
\end{icmlauthorlist}

\icmlaffiliation{sch}{University of California Los Angeles}
\icmlaffiliation{comp}{AWS AI Labs}

\icmlcorrespondingauthor{Wasi Uddin Ahmad}{wasiahmad@ucla.edu}

\vskip 0.3in
]



\printAffiliationsAndNotice{\textsuperscript{*}This work was done during an internship at AWS AI Labs.}  

\begin{abstract}

Recent advances in retrieval-augmented generation (RAG) have initiated a new era in repository-level code completion. However, the invariable use of retrieval in existing methods exposes issues in both efficiency and robustness, with a large proportion of the retrieved contexts proving unhelpful or harmful to code language models (code LMs). In this paper, we propose a selective RAG framework to avoid retrieval when unnecessary. To power this framework, we design a self-supervised learning approach to enable a code LM to accurately self-evaluate whether retrieval can improve its output quality and robustly leverage the potentially noisy retrieved contexts. Using this LM as both the selective RAG policy and the generation model, our framework achieves state-of-the-art repository-level code completion performance on diverse benchmarks including RepoEval, CrossCodeEval, and CrossCodeLongEval, a new long-form code completion benchmark. Meanwhile, our analyses show that selectively retrieving brings as much as 70\% inference speedup in the online serving setting without harming the performance. We further demonstrate that our framework is able to accommodate different generation models, retrievers, and programming languages. These advancements position our framework as an important step towards more accurate and efficient repository-level code completion.

\end{abstract}


\input{text/1_introduction}
\input{text/2_related_work}
\input{text/3_approach}

\input{text/4_experiment}
\input{text/5_results}
\input{text/6_analysis}
\input{text/7_conclusion}


\bibliography{anthology,example_paper}
\bibliographystyle{icml2024}

\include{text/appendix}


\end{document}

%% file: text/1_introduction.tex
\section{Introduction}
\label{section-introduction}

\begin{figure*}[ht!]
\centering
\includegraphics[width=\textwidth]{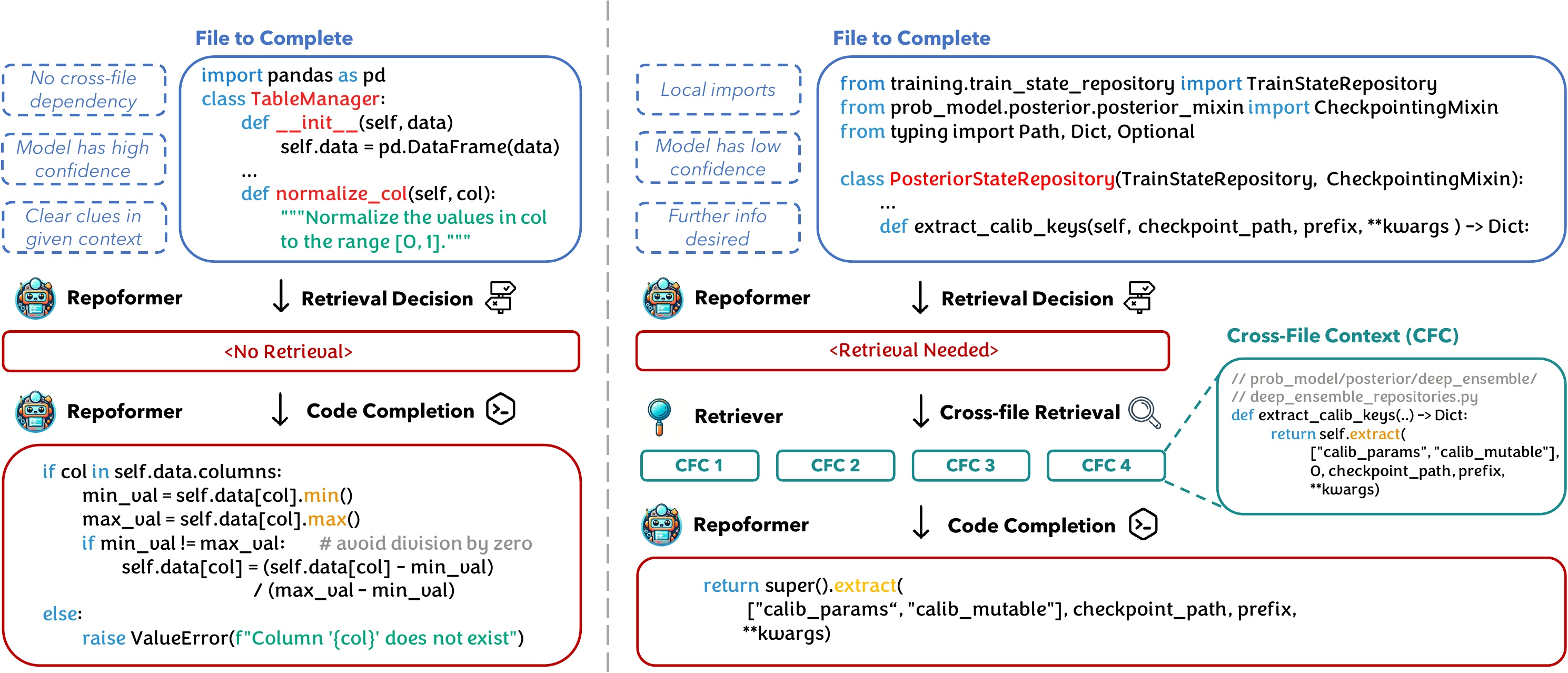}
\caption{An overview of the proposed selective RAG framework. Given the current file context, the system first assesses whether retrieval is required and triggers the retriever if the question can likely be benefited from retrieval (right), abstaining from retrieval otherwise (left). Then, the code LM generates with optional retrieved contexts. With \textsc{Repoformer}, the two stages are streamlined via self-assessment.}
\label{main-framework}
\end{figure*}

Automatic code completion has attracted long-lasting research efforts due to its high practical value in improving programmer productivity \citep{ye2002supporting, hill2004automatic, hellendoorn2017deep}. One particularly challenging scenario is \textit{repository-level code completion}, where a system is required to complete lines, API invocations, or functions in a file from user repositories. For this task, language models for code (code LMs) have emerged as a promising solution due to their ability to leverage the context of the current file to generate coherent code of flexible granularity \citep{tu2014localness, svyatkovskiy2020intellicode, chen2021evaluating}. However, these approaches fail to capture the holistic repository knowledge spanning beyond the current file, such as user-defined APIs and inter-module dependencies \citep{zan-etal-2022-language, zhang2023repocoder,ding2023crosscodeeval}. Recently, the \textit{retrieval-augmented generation} (RAG) paradigm was proposed to bridge the gap: cross-file contexts such as relevant code snippets or documentations are retrieved and provided to code LMs as augmentations to the current file. This approach has shown strong empirical performance and was further advanced by recent literature through designing better retrieval mechanisms for prompting black-box code LMs \citep{lu-etal-2022-reacc, shrivastava2023repository, zhang2023repocoder} and adapting the LM to better leverage structured retrieved contexts such as classes, functions, or APIs \citep{ding2022cocomic, zan-etal-2022-language}.

Despite their encouraging performance, existing RAG-based approaches largely ignore to address a critical question:
\begin{center}
    \textit{Should we always perform retrieval augmentation}? 
\end{center}
Our findings suggest that the answer is predominantly negative. First, in various code completion tasks, we discover that up to 80\% of the retrievals performed by a standard RAG method \textit{do not enhance the performance} of common code LMs such as CodeGen \citep{Nijkamp2022CodeGenAO} and StarCoder \citep{li2023starcoder}, and many degrade the performance by introducing irrelevant information (\cref{80-20-rule}). Second, always retrieving introduces notable \textit{inefficiencies}. For moderately sized repositories, sparse retrieval is already as time consuming as code completion with a 3B code LM (\cref{results-latency} and \cref{section-analysis}). This inefficiency is more pronounced with dense retrieval, enterprise-scale repositories, and iterative RAG methods such as \citet{zhang2023repocoder}.

In this paper, we challenge the assumption of always retrieving by proposing a novel repository-level code completion framework underpinned by a \textit{selective retrieval} mechanism: the system proactively abstains from performing unnecessary or potentially detrimental retrievals (\cref{main-framework} (a)). At the core of our framework is \textsc{Repoformer}, an intelligent code LM fine-tuned for \textit{robust code completion with self-triggered retrieval augmentation}. \textsc{Repoformer} reflects three core principles:

\begin{compactenum}
    \item \textbf{Performance-oriented self-evaluation.} After observing the current file, \textsc{Repoformer} explicitly expresses the likelihood that its prediction quality could be improved by cross-file retrieval. Our training strategy enables the model to combine two factors in this decision: the \textit{code LM} already knowing the answer without retrieval \citep{kadavath2022language} and the \textit{code completion question} not depending on cross-file information and thus retrieval is likely uninformative. 
    \item \textbf{Robustness to retrieved contexts.} 
    \textsc{Repoformer} learns to use the retrieved contexts to improve the quality of its output and avoid performance drops caused by potentially noisy retrieved information.
    \item \textbf{Generalizability.} The aforementioned two abilities must generalize to any completion granularity, programming language, and retriever choice. In addition, \textsc{Repoformer} should be able to function as a plug-and-play selective retrieval policy when other models are employed as the generation model. 
\end{compactenum}

We posit that these abilities can be faithfully obtained by learning from \textit{simulations of RAG}. Specifically, we leverage a large number of permissively licensed repositories, sample diverse blanks to complete, and pair them with the retrieved repository-level cross-file contexts. Then, for a given code LM, the ground-truth label for selective retrieval is obtained by contrasting the quality of its outputs with and without retrieval augmentation. With this dataset, we design a self-supervised objective to jointly train code LMs to accurately self-evaluate the need for retrieval and robustly complete the code with the optional retrieval augmentation (\cref{section-repoformer-method}). 

We perform comprehensive evaluations on a range of repository-level code completion tasks from RepoEval \citep{zhang2023repocoder}, CrossCodeEval \citep{ding2023crosscodeeval}, and CrossCodeLongEval a new large-scale benchmark focusing on code chunk and function completion. Results show that \textsc{Repoformer} achieves strong performance, outperforming always retrieving with the same-sized StarCoderBase by more than 3 absolute points for edit similarity across multiple tasks. The 3B \textsc{Repoformer} performs on par with always retrieving using the 16B StarCoder, and the 16B \textsc{Repoformer} achieves state-of-the-art performance across all the tasks (\cref{section-result-performance}). Furthermore, our framework allows for up to 70\% inference speedup without harming accuracy. We also establish that \textsc{Repoformer} can accelerate RAG with larger black-box LMs as a plug-and-play selective RAG policy, improving the performance while reducing the latency of line and API completion to 75\% (\cref{results-latency}). 

Finally, in \cref{section-analysis}, we provide comprehensive analyses on \textsc{Repoformer}'s generalization ability. We show that \textsc{Repoformer} makes precise retrieval abstention decisions, is robust to retrieved contexts, and performs well when tested in other languages or with other retrievers. To facilitate future research on repository-level code completion, we will release our implementation and the CrossCodeLongEval benchmark at \url{https://repoformer.github.io/}. 

%% file: text/2_related_work.tex
\section{Related Work}
\paragraph{Repository-level Code Completion} Accurately completing the code in repositories has been a challenging research problem due to cross-file dependency patterns caused by modular design \citep{parnas1972criteria, tu2014localness}. Early works propose application-specific training methods for n-gram LMs \citep{tu2014localness}, RNNs \citep{hellendoorn2017deep,wang2021cocosum}, and Transformers \citep{svyatkovskiy2020intellicode} to leverage structured knowledge beyond current file's context. Recent studies investigate fine-tuning powerful pre-trained code LMs \citep{chen2021evaluating, Nijkamp2022CodeGenAO, li2023starcoder} to better leverage retrieved knowledge provided in context such as code and documentation snippets \citep{zan-etal-2022-language, ding2022cocomic, shrivastava2023repofusion}. Concurrently, other studies show that black-box code LMs can already take advantage of in-context knowledge, depending on how well the knowledge is retrieved and formatted \citep{lu-etal-2022-reacc, zhou23docprompting, shrivastava2023repository, zhang2023repocoder}. This approach does not require one to train the LM and thus promises better generalization. Orthogonal to these studies, this paper identifies and addresses the robustness and efficiency issues caused by invariably performing the retrieval augmentation. Our solution takes the form of selective retrieval augmentation through self-assessment.

\paragraph{Adaptive RAG} This paper is consistent with the recent trend of making the RAG paradigm active and adaptive. A core question is finding an effective policy to decide \textit{when to retrieve}. \citet{he-etal-2021-efficient} propose to learn to adjust the importance weight of retrieval based on language modeling performance. \citet{drozdov-etal-2022-cant} proposes to upweight the retrieved information when the retrieval has high quality. \citet{li-etal-2023-web} and \citet{jiang2023active} suggest that retrieval should be performed only when LMs have a high predictive uncertainty. \citet{mallen-etal-2023-trust} discover that retrieval can be avoided for popular facts. Concurrent to this work, two new studies approach adaptive RAG from a learning perspective. SKR \citep{wang2023self} collects instances where retrieval is not helpful for black-box LMs and proposes several methods to predict these instances. Self-RAG \citep{asai2023self} utilizes GPT-4 \citep{OpenAI2023GPT4TR} as a knowledge engine to distill a smaller LM to evaluate whether answering a question can be benefited from retrieval. In comparison, this paper highlights the importance of understanding whether an LM knows the answer \citep{kadavath2022language} in forming the retrieval policy. We introduce a simple yet effective scheme to fine-tune a code LM for faithful self-evaluation without extra modules (SKR), knowledge store (SKR), or labels generated by an oracle LM (Self-RAG). We show that our approach leads to no performance harms (\cref{section-result-performance}), substantial speedup (\cref{results-latency}), and a high decision accuracy (\cref{section-analysis}). 

%% file: text/3_approach.tex
\section{Approach}

In this section, we first briefly formulate the repository-level code completion task and the considered RAG setup. Then, we illustrate the details of the proposed framework. 

\subsection{Background}
\paragraph{Problem Formulation}
We denote each \textit{repository-level code completion} task as $(X_l, X_r, Y, F)$. $Y$ is the ground truth completion that needs to be generated. In this paper, $Y$ always contains one or more consecutive lines of code. $X_l$ and $X_r$ are the code to the left/right of $Y$ in the same file. We will use the left/right context to refer to them. $F$ is the set of other files in the repository. A code completion system utilizes $X_l$, $X_r$, and $F$ to generate a hypothesis $\hat{Y}$. 

\paragraph{Retrieval-Augmented Generation} We follow the RG-1 formulation in \citet{zhang2023repocoder} to execute RAG for code completion in four stages: \textit{indexing}, \textit{query formation}, \textit{retrieval}, and \textit{generation}. We consider two components:

\begin{compactitem}
    \item An \textbf{in-repository retriever $\mathcal{R}$} that queries $F$ with information from $X_l$ and $X_r$ and returns relevant cross-file contexts $CC$. $CC$ consists of $k$ code chunks $cc_1, cc_2, ..., cc_k$, each of which contains consecutive lines of code extracted from a file in $F$. We mainly use Jaccard similarity \citep{jaccard1912distribution} as $\mathcal{R}$ due to its speed and strong performance \citep{zhang2023repocoder}.
    \item A \textbf{code LM $\mathcal{M}$} that leverages $X_l$, $X_r$, and $CC$ to output $\hat{Y}$. The inclusion of $X_r$ and $CC$ is optional. In this paper, we always directly provide $X_r$ in the prompt in addition to $X_l$ \citep{shrivastava2023repository,Pei_Zhao_Lausen_Zha_Karypis_2023}. We provide empirical support for this design in \cref{section-why-infilling}.
\end{compactitem}

Full documentation of the RAG stages and their hyperparameters are provided in \cref{detailed-rag-setup} for further reference. 

\subsection{Self-selective RAG for Code Completion}

Central to our framework is the idea of \textit{selective RAG}, where the system decides whether the LM's generation could benefit from retrieved contexts and abstains from retrieval augmentation when it is deemed unnecessary (\cref{main-framework}).

For this selective decision, two traditional heuristics are relevant: (1) performing a \textit{trial retrieval} and only augmenting the high-relevance contexts (e.g., \citet{drozdov-etal-2022-cant}) or (2) performing a \textit{trial generation} and conducting RAG only when the model's uncertainty is high (e.g., \citet{jiang2023active}). For repository-level code completion, these strategies are informative to some extent: in line completion and API completion from RepoEval, both heuristics can maintain the same level of performance with only 50\% retrieval budget. However, we find that they fail to generalize well to all tasks and still incur a high latency cost as they need to conduct retrieval to make the decisions (\cref{appendix-retriever-for-srag}). 

Instead, our framework adopts a \textit{self-selective RAG} formulation. After observing $X_l$ and $X_r$, the LM directly self-triggers cross-file retrieval by generating a special token \texttt{\textcolor{blue}{\textless cc\textgreater}} or abstains from retrieval via an empty token $\phi$\footnote{In practice, instead of greedily decoding \texttt{\textcolor{blue}{\textless cc\textgreater}}, we check whether its probability exceeds a certain threshold.}. This approach is inspired by the explorations in \citet{kadavath2022language}, which show that an LM can be trained to predict whether it knows the answer or not without retrieval. Beyond this \textit{self-knowledge}, our model also combines the \textit{question's characteristics} (i.e., whether retrieving cross-file information can likely help or not) in its judgment, as we will discuss in the next section. Finally, after the optional retrieval, the LM proceeds with the code completion with $X_l$, $X_r$, combined with $CC$ if retrieval is triggered. 

Implementation-wise, self-selective RAG's inference is conveniently modeled as an extension to fill-in-the-middle \citep{bavarian2022efficient}, with the entire process executed in a single left-to-right pass (\cref{self-selective-rag-prompt}). One advantage of this design is the \textit{flexibility}. The LM possesses the ability for RAG and fill-in-the-middle, and can seamlessly self-switch between the two when encountering different questions. Users can also easily adjust the ratio between the two through the retrieval threshold. Another advantage is its \textit{efficiency}. The selective decision overhead is only a single forward pass, a significant save compared to making the retrieval decision via trial generation or trial retrieval. When the LM abstains from retrieval, it can directly proceed with generation and the retrieval overhead is completely avoided.

\subsection{Self-supervised Multi-task Learning}
\label{section-repoformer-method}

To power self-selective RAG, the LM needs two crucial abilities: accurate self-assessment and robustness to the retrieved context. We design a contrastive data labeling scheme to mine self-supervision from public repositories, followed by fine-tuning with a novel multi-task objective. 

\paragraph{Data construction} We leverage large-scale permissively licensed repositories from the Stack \citep{kocetkov2022stack} and create the fine-tuning data via a three-step procedure:

\begin{compactenum}
    \item \textbf{Sample} target lines $Y$ that are either (1) random code chunks of varied lengths or (2) function bodies. 
    \item \textbf{Retrieve} $CC$ using the current file. We include $Y$ in the query for 50\% of the data\footnote{The main goal of the design is to align better with both non-iterative and iterative RAG use cases. During testing, a user may retrieve with both the in-file context and $Y’$, a model’s draft prediction, which results in a $CC$ distribution close to that with $Y$ in the query \citep{zhang2023repocoder}. }.
    \item \textbf{Label} whether extending the current file with $CC$ can improve a code LM $\mathcal{M}$'s code completion quality by more than a threshold $T$, measured by Edit Similarity (ES, definition in \cref{repoformer-implementation-details}) against $Y$. 
\end{compactenum}

\begin{figure}[t]
\centering
\includegraphics[width=0.97\linewidth]{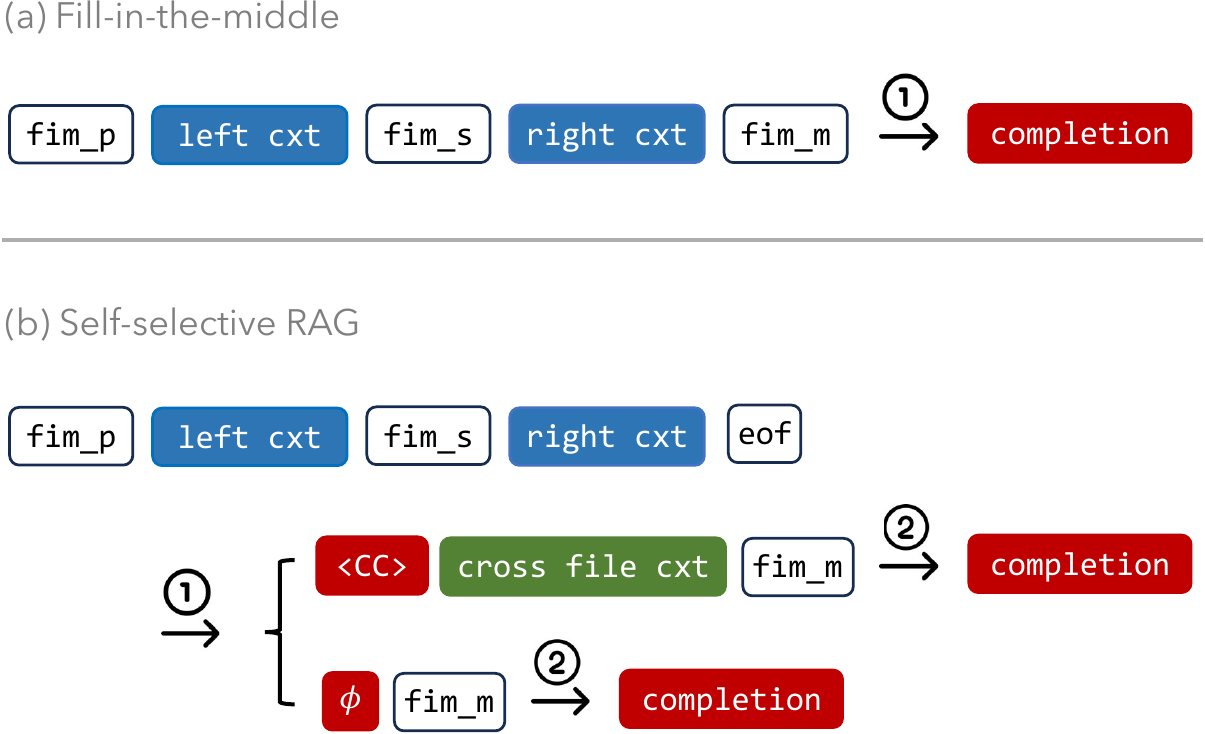}
\caption{A comparison between fill-in-the-middle and self-selective RAG. We mark the end of the current file with a new token \texttt{\textcolor{blue}{\textless eof\textgreater}}, which triggers the LM's self-evaluation. $\rightarrow$ denotes the invocation of the LM. We color current-file context, retrieved contexts, and LM-generated parts in blue, green, and red respectively. \texttt{fim\_p}, \texttt{fim\_s}, and \texttt{fim\_m} refer to the special tokens for fill-in-the-middle: \texttt{fim\_prefix}, \texttt{fim\_suffix}, and \texttt{fim\_middle}. These tokens are already learned during the pre-training. }
\label{self-selective-rag-prompt}
\vspace{-2mm}
\end{figure}

The full algorithms are presented in \cref{appendix-data-creation-algo}. After running the algorithm, we obtain the fine-tuning instances, each in the form $(X_l, X_r,\ Y,\ CC,\ label)$.

\paragraph{Verbalization} Each instance is verbalized into a sequence for fine-tuning. If $label$ is false, only $X_l$ and $X_r$ are provided preceding $Y$. Otherwise, we additionally provide $CC$ after the special token \texttt{\textcolor{blue}{\textless cc\textgreater}}. The two verbalizations correspond to the two branches in \cref{self-selective-rag-prompt} (b).

\paragraph{Training Objective} We introduce two losses, $\mathcal{L}_{eval}$ for self-assessment and $\mathcal{L}_{gen}$ for code generation.
\begin{enumerate}
    \item $\mathcal{L}_{eval}$: a cross-entropy loss on predicting \texttt{\textcolor{blue}{\textless cc\textgreater}} immediately following \texttt{\textcolor{blue}{\textless eof\textgreater}}.  
    \begin{equation}
    \mathcal{L}_{eval}=-\log p_\mathcal{M}(\text{\texttt{\textless cc\textgreater}}|X_l,X_r)
    \end{equation}
    \item $\mathcal{L}_{gen}$: a cross-entropy loss on the tokens following \texttt{\textcolor{blue}{\textless fim\_middle\textgreater}}. Depending on $label$, $\mathcal{L}_{gen}$ represents either code completion with only in-file information or retrieval-augmented code completion. 
    \begin{equation}
    \mathcal{L}_{gen} = \begin{cases}
        -\log p_\mathcal{M}(Y|X_l,X_r,CC),& \text{if } label\\
        -\log p_\mathcal{M}(Y|X_l,X_r),              & \text{otherwise}
    \end{cases}
    \end{equation}
\end{enumerate}
\vspace{-2mm}
The final training objective is $\lambda\mathcal{L}_{eval}+\mathcal{L}_{gen}$, a weighted combination of the two losses. We do not supervise the model on predicting the other tokens in $X_l$, $X_r$, $CC$, or the special tokens for fill-in-the-middle. Teacher forcing is used just as in normal causal language model training.

%% file: text/4_experiment.tex
\section{Experimental Setup}
\label{section-benchmarking-setup}

\subsection{\textsc{Repoformer} Implementation Details}
\label{repoformer-implementation-details}

\paragraph{Training Data} We sample Python repositories from the Stack \citep{kocetkov2022stack}. Basic filtering are applied to retain 18k repositories that have (1) at least five Python files, (2) at least three imports per file, and (3) at least two local imports per file. These criteria ensure the existence of local dependencies where RAG could be helpful. We use $\mathcal{M}$ = StarCoderBase-1B and $T$ = $0$ to label 240k chunk and 120k function completion instances. We reserve 500 repositories for validation and use the rest for training.

\paragraph{Training} We fine-tune the 1B, 3B, 7B, and 16B variants of StarCoderBase with $\lambda=1.0$, maximum sequence length 2048, learning rate 2e-5, batch size 512, 50 warmup steps, and a linear learning rate decay. The models are trained for 2 epochs, which approximately takes 8, 12, 20, and 50 hours for the 1B/3B/7B/16B models respectively with 8 Nvidia A100 GPUs (40G memory). Our implementation is based on \citet{jain-etal-2023-contraclm}\footnote{\small \url{https://github.com/amazon-science/ContraCLM}}. 
 We will call our models \textsc{Repoformer-1B/3B/7B/16B}. We have also applied the same method to train a multilingual version of \textsc{Repoformer} on a mixture of Python, Java, C\#, and Typescript repositories. As we focus on the methodological discussion in the main text, we refer interested readers to \cref{appendix-multilingual-results} for the detailed experiment setup and results.
 
\paragraph{Hyperparameter optimization} We conduct a grid search with StarCoderBase-1B on the following search space: learning rate \{1e-5, 2e-5, 5e-5\}, $\lambda$ \{0.2, 1.0, 2.0, 5.0\}, training epochs \{1, 2, 5\}, and warmup steps \{50, 100\}. The best hyperparameters are selected based on the code completion performance on the validation dataset. 

\subsection{Evaluation Setup}

\paragraph{Evaluation Datasets} We evaluate on RepoEval \citep{zhang2023repocoder}, which consists of line, API, and function completion tasks created from 32 Python repositories. To investigate the generalization to other languages, we also evaluated the original CrossCodeEval \citep{ding2023crosscodeeval}, which features line completion instances covering four languages: Python, Java, C\#, and TypeScript (\cref{appendux-crosscodeeval-results}). Observing that RepoEval has a limited repositrory coverage and that CrossCodeEval has a limited task coverage, we additionally leverage 1500 raw Python repositories from CrossCodeEval to create a new chunk and function completion benchmark, which we call CrossCodeLongEval. We detail the dataset creation process and basic statistics in \cref{appendix-cceval-creation}. For the rest of this paper, we will use CCEval to refer to both CrossCodeEval and CrossCodeLongEval interchangeably, and use the specific language and task (line, chunk, or function completion) to differentiate them. 

\begin{figure*}[ht!]
\centering
\includegraphics[width=\textwidth]{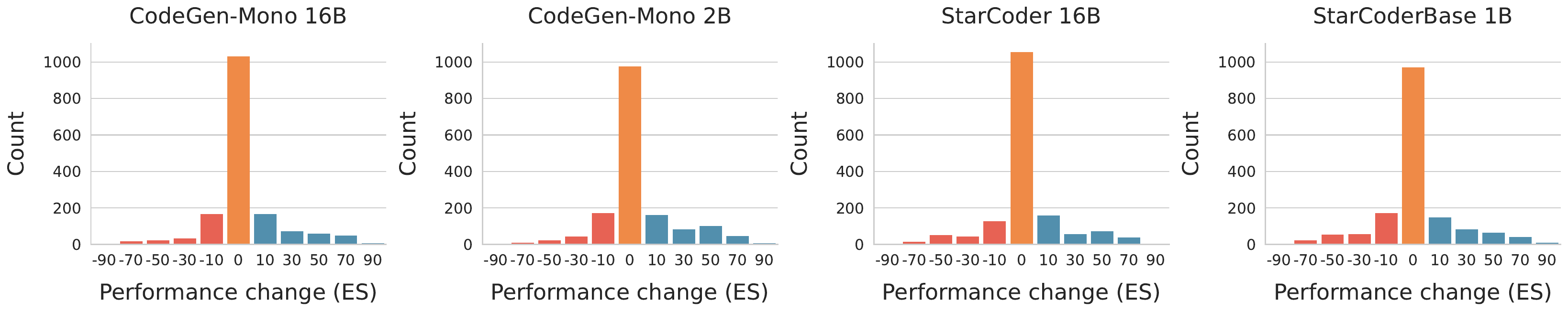}
\caption{The performance gain on RepoEval API completion from retrieved cross-file contexts. Each bucket contains values ranging from label-10 to label+10 except for the central bucket, which corresponds to exactly 0. The retrieved contexts only improve the performance in about 20\% of instances. The trend is consistent across al\textbf{}l the evaluated LM families and sizes.}
\vspace{2mm}
\label{retrieval_effectiveness_api}
\end{figure*}

\paragraph{Evaluation Metrics} We evaluate $\hat{Y}$ with both \textit{reference-based} and \textit{execution-based} evaluation. For reference-based evaluation, exact match (EM) and edit similarity (ES) are reported. Following \citet{zhang2023repocoder}, ES is defined as 
\begin{equation}
ES(\hat{Y}, Y) = \frac{1-Lev(\hat{Y}, Y)}{\max(|\hat{Y}|, |Y|)},
\end{equation}
where $Lev$ is the Levenshtein distance \citep{levenshtein1966binary}. We report $ES\times100$ in all the tables following \citet{zhang2023repocoder} for better readability. For execution-based evaluation, we report the unit test pass rate (UT). $\hat{Y}$ is said to pass the unit tests if replacing $Y$ with $\hat{Y}$ does not cause any unit test to fail. We implement simple post-processing procedures to handle common cases such as excessive lines in model's outputs, which are documented in \cref{detailed-rag-setup}.

\paragraph{Models} We experiment on two families of strong code LMs.  \textbf{CodeGen-Mono} \citep{Nijkamp2022CodeGenAO} is pretrained sequentially in natural language, multilingual code, and a Python corpus. \textbf{StarCoder} and \textbf{StarCoderBase} \citep{li2023starcoder} are trained with fill-in-the-middle ability on a large corpus of multilingual code, GitHub issues, Git commits, and Jupyter notebooks. StarCoder is obtained by training StarCoderBase on an additional Python corpus. 

%% file: text/5_results.tex
\section{Results}

\subsection{Is retrieval always helpful?}
\label{80-20-rule}

As a proof of concept, we first show that on a range of repository-level code completion tasks, the retrieved contexts often fail to improve code LMs' generation quality. 

In \cref{tab:retrieval_effectiveness_func} and \cref{retrieval_effectiveness_api}, we evaluate four code LMs on function completion and API completion from RepoEval. For each model, we report the instance-level performance change from code completion only using $X_l$ and $X_r$ to retrieval-augmented code completion with $X_l$, $X_r$, and $CC$ (detailed prompts in \cref{detailed-rag-setup}). 

The results reveal an intriguing pattern: for repository-level code completion, the help from cross retrieval is often sparse. Specifically, retrieval improves LMs' performance on only 20\% or fewer instances. For more than 60\% of the instances, retrieval augmentation does not affect the performance at all\footnote{Upon a manual inspection, we find that most of the outputs in this category are also not changed by retrieval at all.}. Finally, another 20\% retrievals actually harm the performance, almost as often as the first case. The observed trends are consistent for both API and function completion and hold for both small-sized (1B and 2B) and moderate-to-large (around 16B) code LMs. The generality of this observation is further confirmed by an analysis of \textsc{Repoformer}'s training data, where we find that retrieval improves the performance for only fewer than 30\% instances (\cref{appendix-training-data-analysis}). Together, these findings highlight the suboptimality of the always retrieving and augmenting the cross-file contexts and thus motivate our selective retrieval proposal.  

\subsection{\textsc{Repoformer} achieves strong code completion performance via selective RAG}
\label{section-result-performance}

\input{tables/retrieval_effectiveness_func}

\input{tables/main_eval_v2}

Next, we evaluate the code completion performance of \textsc{Repoformer}. We compare the following three settings\footnote{We do not consider iterative retrieval because we find that single-iteration RAG already achieves the majority of the performance gains from multi-iteration RAG.}. For the first two baselines, we use the state-of-the-art single-iteration prompting pipeline (\citet{zhang2023repocoder}, detailed in \cref{detailed-rag-setup}). We use StarCoder models due to their strong performance among the open-source code LMs. 

\begin{compactenum}
    \item \textbf{No Retrieval}. This baseline only provides $X_l$ and $X_r$ to the model in the prompt.
    \item \textbf{Always Retrieving}. This baseline always augments $X_l$ and $X_r$ with the retrieved $CC$.
    \item \textbf{Selective Retrieval}. We provide \textsc{Repoformer} with $X_l$ and $X_r$ in the prompt, optionally augmented with $CC$ based on two selective RAG policies:
        \begin{compactitem}
            \item \textbf{Greedy Selection}. Retrieval is performed if \texttt{\textless cc\textgreater} is the most likely token following \texttt{\textless eof\textgreater}. 
            \item \textbf{Threshold Selection}. If the probability of \texttt{\textless cc\textgreater} following \texttt{\textless eof\textgreater} is greater than a threshold $T$, retrieval augmentation is performed\footnote{We find that $T=0.15$ for function completion and $T=0.2$ for the other tasks generally work well. These two thresholds are always used unless otherwise stated.}.
        \end{compactitem}
\end{compactenum}

The results are summarized in \cref{tab:main_eval}. Compared to no retrieval and always retrieving with StarCoderBase of the same size, \textsc{Repoformer}'s selective retrieval strategy exhibits strong performance improvements across all the tasks and both lexical-based and execution-based metrics. Via the threshold selection strategy, \textsc{Repoformer-3B} can outperform StarCoderBase-7B on most of the tasks and metrics except EM for API completion, even outperforming the 5x larger StarCoder in terms of ES for API and chunk completion. Finally, The \textsc{Repoformer-16B} model outperforms the strongest StarCoder baseline by 3\%, averaged across all tasks, setting up the new start-of-the-art for repository-level code completion. We also experimentally confirm that the performance improvement from our framework can generalize to three \textit{languages beyond Python} (\cref{appendux-crosscodeeval-results}) as well as \textit{dense retrieval} instead of Jaccard similarity (\cref{appendix-retriever-robustness}). In later sections, we demonstrate that the observed success is due to both the ability to accurately abstain from retrieval and the improved robustness to retrieval.

In terms of code completion accuracy, the threshold selection strategy outperforms the greedy selection strategy on all the tasks. In the next section, we show that the two strategies represent different ways to achieve a good balance between accuracy and inference budget.

\subsection{\textsc{Repoformer} improves inference efficiency}
\label{results-latency}

We illustrate the benefits of \textsc{Repoformer} for saving the inference latency in a realistic ``online serving" setting.

\paragraph{Latency Model} We assume that indexing has already been done for the working repository. Given a code completion request containing the current file $(X_l, X_r)$, the system issues three processes at the same time:
\begin{compactitem}
    \item $P_1$: make a retrieval decision using \textsc{Repoformer}.
    \item $P_2$: use a code LM $\mathcal{M}$ to generate $\hat{Y}$ without $CC$.
    \item $P_3$: retrieve $CC$ and generate $\hat{Y}$ with $CC$ using $\mathcal{M}$.
\end{compactitem}

Depending on the result of $P_1$, the system waits for either $P_2$ or $P_3$ and ignores the other process. If $\mathcal{M}$ is \textsc{Repoformer}, $P_1$ can be merged with $P_2$ by forcing $\mathcal{M}$ to generate a hypothesis without $CC$ after collecting the retrieval decision. We consider three latency terms: (1) $T_{d}$, time required for the retrieval decision, (2) $T_{r}$, the retrieval latency, and (3) $T_{g}$, the generation latency. Then, the latency for $P_1$, $P_2$, and $P_3$ are $T_{d}$, $T_{g}$, and $T_{r}+T_{g}$. When $\mathcal{M}$ is \textsc{Repoformer} or a model larger than \textsc{Repoformer}, we have $T_{d} < T_{g} < T_{r}+T_{g}$. Therefore, the latency of the entire system is $T_{g}$ or $T_{r}+T_{g}$ depending on $P_1$. Using this latency model, we benchmark the latency of various selective retrieval settings on RepoEval with the vllm library \citep{kwon2023efficient} on a single Nvidia A100 GPU (80G).

\input{tables/latency_self}

First, we consider $\mathcal{M}$ = \textsc{Repoformer} and present the results in \cref{latency-repoformer}. Line and API completion are presented to cover short and moderate target lengths\footnote{We omit the function completion results as RepoEval uses very small repositories for function completion for easier unit testing.}. Both selective strategies significantly improve the latency, with a different trade-off: threshold selection results in \textit{improvements for both accuracy and latency} compared to always retrieving, while using greedy selection results in a larger latency gain with a minor performance degradation (around 1.0 ES). It is worth mentioning that the latency improvement from selective RAG could be further enhanced with a more advanced retrieval setup. For instance, conducting dense retrieval on large repositories often consumes more than 80\% of the entire RAG pipeline's latency. Then, a 20\% RAG policy could translate into more than 70\% speedup. We empirically verify this statement in \cref{appendix-retriever-robustness}. 

Next, we consider using diverse larger LMs as $\mathcal{M}$ in the code completion framework and using selection\textsubscript{T} with \textsc{Repoformer-1B} as a \textit{plug-and-play} selective RAG policy to decide whether retrieval should be performed. We experiment on a diverse set of LMs: StarCoderBase, Code Llama \citep{roziere2023code}\footnote{We accessed the model through Amazon SageMaker (\url{https://docs.aws.amazon.com/sagemaker/}).}, CodeGen25 \citep{Nijkamp2023codegen2}, and ChatGPT\footnote{We use \texttt{gpt-3.5-turbo-0613} via the OpenAI API.}. As shown in \cref{latency-transfer}, the selective predictions from \textsc{Repoformer-1B} successfully reduce the inference latency with different larger LMs by approximately 25\% \textit{}\textit{while improving their accuracy}. Collectively, the findings indicate that \textsc{Repoformer} has acquired robust selective retrieval capabilities that could generalize to diverse types of code LMs.

\input{tables/latency_transfer}

%% file: tables/retrieval_effectiveness_func.tex
\begin{table}[t!]
    \centering
    \resizebox{\linewidth}{!} {%
    \begin{tabular}{c | c | c c | c c c}
        \hline
        \multirow{2}{*}{Model} & \multirow{2}{*}{Size} &  \multicolumn{2}{c|}{Performance (UT)}  &   \multicolumn{3}{c}{UT Change} \\
        & & $X_l$ + $X_r$ & $X_l$ + $X_r$ + $CC$ & $\downarrow$ & $=$ & $\uparrow$  \\
        \hline
        CodeGen-Mono & 16B & 23.74 & 24.18 & 23 & 407 & 25 \\
        CodeGen-Mono &  2B & 30.55 & 32.51 & 18 & 400 & 37 \\
        StarCoder & 16B & 34.73 & 42.86 & 16 & 386 & 53 \\
        StarCoderBase & 1B & 22.20  & 25.71 & 16 & 407 & 32 \\
        \hline
    \end{tabular}
    }
    \caption{The performance change on RepoEval function completion exhibited by four models from retrieved cross-file contexts. For the majority of the instances, RAG does not improve the performance. ``$\uparrow$", ``$=$", ``$\downarrow$" denote the counts for performance increase, no performance change, and performance drop.}
    \label{tab:retrieval_effectiveness_func}
\end{table}

%% file: tables/main_eval_v2.tex
\begin{table*}[ht!]
 \centering
 \resizebox{\linewidth}{!} {%
\begin{tabular}{l|c|c|cc|cc|cc|cc|c}
\hline
&  &  & \multicolumn{6}{c|}{\textbf{RepoEval}} & \multicolumn{3}{c}{\textbf{CrossCodeLongEval}}  \\ 
&  &  & \multicolumn{2}{c}{\textbf{(Line)}}  & \multicolumn{2}{c}{\textbf{(API)}}   & \multicolumn{2}{c|}{\textbf{(Function)}} & \multicolumn{2}{c}{\textbf{(Chunk)}} & \textbf{(Function)}  \\
\multicolumn{1}{c|}{\multirow{-3}{*}{\textbf{Size}}} & \multicolumn{1}{c|}{\multirow{-3}{*}{\textbf{Model}}} & \multicolumn{1}{c|}{\multirow{-3}{*}{\textbf{RAG Policy}}} & \textbf{EM} & \multicolumn{1}{c}{\textbf{ES}} & \textbf{EM} & \multicolumn{1}{c}{\textbf{ES}} & \textbf{UT} & \textbf{ES} & \textbf{EM} & \multicolumn{1}{c}{\textbf{ES}} & \textbf{ES} \\
\hline
\hline
\multirow{4}{*}{1B} & \multirow{2}{*}{\textsc{StarCoderBase}} & No  & 43.44 & 67.77 & 37.81 & 66.54 & 22.20 & 47.65 & 31.08 & 60.09 & 47.49 \\
 & & Always  &  51.19 & 72.30 & 43.94 & 69.17 & 25.71 & 55.64 & 37.22 & 63.73 & 50.50 \\
 & \multirow{2}{*}{\textsc{Repoformer}} & Selective\textsubscript{G} &  51.90 & 74.50 & 43.50 & 71.00 & 24.00 & 53.10 & 38.52 & 68.08 & 52.09 \\
 & & Selective\textsubscript{T} & \textbf{54.40} & \textbf{76.00} & \textbf{46.10} & \textbf{72.70} & \textbf{28.79} & \textbf{57.30} & \textbf{41.92} & \textbf{69.97} & \textbf{53.71 }\\
\hline
\hline
\multirow{4}{*}{3B} & \multirow{2}{*}{\textsc{StarCoderBase}} & No & 49.00 & 72.12 & 40.44 & 69.02 & 24.84 & 51.22 & 36.14 & 64.65 & 49.88 \\
 & & Always  & 56.69 & 76.68 & 47.00 & 72.62 & 29.67 & 57.68 & 42.26 & 67.74 & 53.39 \\
 & \multirow{2}{*}{\textsc{Repoformer}} & Selective\textsubscript{G} & 56.30 & 77.60 & 46.10 & 73.60 & 28.57 & 54.70 & 42.06 & 70.70 & 54.47 \\
 & & Selective\textsubscript{T} & \textbf{59.63} & \textbf{79.02} & \textbf{49.31} & \textbf{74.96} & \textbf{32.96} & \textbf{60.56} & \textbf{46.66} & \textbf{72.23} & \textbf{56.24} \\
 \hline
\hline
\multirow{4}{*}{7B} & \multirow{2}{*}{\textsc{StarCoderBase}} & No & 51.88 & 74.03 & 43.31 & 70.79 & 25.49 & 52.28 & 38.88 & 66.61 & 52.45 \\
 & & Always  & 59.44 & 78.15 & 49.56 & 73.65 & 31.43 & 58.51 & 44.44 & 69.53 & 55.41 \\
 & \multirow{2}{*}{\textsc{Repoformer}} & Selective\textsubscript{G} & 56.00 & 76.63 & 48.06 & 75.03 & 30.77 & 55.27 & 43.80 & 72.46 & 56.14 \\
 & & Selective\textsubscript{T} & \textbf{59.63} & \textbf{78.63} & \textbf{50.87} & \textbf{76.89} & \textbf{35.16} & \textbf{60.64} & \textbf{46.88} & \textbf{74.20} & \textbf{57.18} \\
 \hline
\hline
\multirow{4}{*}{16B} & \multirow{2}{*}{\textsc{StarCoder}} & No & 55.25 & 76.07 & 44.50 & 71.00 & 34.73 & 53.60 & 42.58 & 69.40 & 54.20 \\
 & & Always &  61.25 & 79.24 & 51.12 & 74.50 & 42.86 & 60.96 & 47.90 & 71.90 & 58.06 \\
 & \multirow{2}{*}{\textsc{Repoformer}} & Selective\textsubscript{G} & 58.13 & 78.81 & 48.69 & 76.23 & 42.42 & 58.42 & 45.00 & 73.36 & 57.71 \\
 & & Selective\textsubscript{T} & \textbf{61.75} & \textbf{80.34} & \textbf{51.88} & \textbf{77.93} & \textbf{44.18} & \textbf{62.58} & \textbf{49.18} & \textbf{75.50} & \textbf{58.93} \\
 \hline
\end{tabular}
}
\caption{ Experiment results on RepoEval and CrossCodeLongEval. The best performance among each model size is boldfaced. We use Selective\textsubscript{G} and Selective\textsubscript{T} to denote the greedy selection and the threshold selection strategy for selective retrieval. \textsc{Repoformer} greatly outperforms \textsc{StarCoderBase} of the same size while consuming a smaller retrieval budget. Among the two selective policies, threshold selection enables the best selective RAG performance.}
\label{tab:main_eval}
\end{table*}

%% file: tables/latency_self.tex
\begin{table}[t]
\resizebox{\linewidth}{!} {
\begin{tabular}{l|c|c@{}c@{}c|c@{}c@{}c}
\hline
& \multirow{2}{*}{\textbf{RAG Policy}} & \multicolumn{3}{c|}{\textbf{API Completion}} & \multicolumn{3}{c}{\textbf{Line Completion}}  \\
& & \textbf{ES} & \textbf{\%RAG} & \textbf{SU} & \textbf{ES} & \textbf{\%RAG} & \textbf{SU} \\
\hline
& Always & 72.02 & $100\%$ & $0\%$ & 75.91 & $100\%$ & 0\% \\
& Selective\textsubscript{G}  & 71.04 & $18\%$ & $69\%$ & 74.50 & $19\%$  & $61\%$ \\
\multirow{-3}{*}{1B} & \multicolumn{1}{c|}{Selective\textsubscript{T}} & 72.72 & $61\%$ & $28\%$ & 76.00 & $62\%$ & $27\%$ \\
\hline
& Always & 74.66 & $100\%$ & $0\%$ & 78.68 & $100\%$ & 0\% \\
& Selective\textsubscript{G} & 73.60 & $19\%$  & $46\%$ & 77.60 & $20\%$ & $43\%$ \\
\multirow{-3}{*}{3B} & \multicolumn{1}{c|}{Selective\textsubscript{T}} & 74.96 & $78\%$ & $17\%$ & 79.02  & $74\%$ & $16\%$  \\
\hline
\end{tabular}
}
\caption{RAG latency of \textsc{Repoformer} with two self-selective RAG paradigms. \textbf{\%RAG} = ratio of instances where RAG is performed. \textbf{SU} = Speedup compared to always retrieving (the higher, the better).  Compared to the always retrieving baseline, the threshold selection strategy consistently demonstrates gains in both accuracy and latency. The greedy selection strategy shows much larger latency gains with a small performance degradation.}
\label{latency-repoformer}
\end{table}

%% file: tables/latency_transfer.tex
\setlength{\tabcolsep}{4pt}
\begin{table}[t]
 \resizebox{1.0\linewidth}{!} {
\begin{tabular}{l|c|cc|cc}
\hline
\multirow{2}{*}{\textbf{Model}} & \multirow{2}{*}{\textbf{RAG Policy}} & \multicolumn{2}{c|}{\textbf{API Completion}} & \multicolumn{2}{c}{\textbf{Line Completion}}  \\
& & \textbf{ES} & \textbf{SU} & \textbf{ES} & \textbf{SU} \\
\hline
& Always Retrieving & 73.65 & $0\%$ & 78.15 & $0\%$ \\
\multirow{-2}{*}{\textsc{SCB-7B}}  & \textsc{Repoformer-1B}  & 74.10 &  $24\%$  & 78.31 & $25\%$ \\
\hline
& Always Retrieving & 74.50 & $0\%$ & 79.24 & $0\%$ \\
\multirow{-2}{*}{\textsc{SCB-16B}} & \textsc{Repoformer-1B}  & 74.84 &  $24\%$  & 79.48 & $24\%$  \\
\hline
& Always Retrieving & 63.07 & 0\% & 68.42 & 0\% \\
\multirow{-2}{*}{\textsc{CG25-7B}} & \textsc{Repoformer-1B}  & 63.37 & 20\% & 68.86 & 29\%  \\
\hline
& Always Retrieving & 58.75 & 0\% & 59.99 & 0\% \\
\multirow{-2}{*}{\textsc{CL-7B}} & \textsc{Repoformer-1B}  & 58.91 & 25\% & 60.47 & 28\%  \\
\hline
& Always Retrieving & 61.08 & 0\% & 61.58 & 0\% \\
\multirow{-2}{*}{\textsc{CL-16B}} & \textsc{Repoformer-1B}  & 62.10 & 32\% & 62.45 & 30\%  \\
\hline
& Always Retrieving & 63.38 & 0\% & 61.76 & 0\% \\
\multirow{-2}{*}{\textsc{ChatGPT}} & \textsc{Repoformer-1B}  &  64.01 & 28\% & 61.92 & 18\%  \\
\hline
\end{tabular}
}
\caption{Accuracy and latency of larger code LMs as the generation model and with \textsc{Repoformer-1B} as the policy model for selective RAG. \textsc{SCB} = StarCoderBase, \textsc{CG25} = CodeGen25, \textsc{CL} = Code Llama. \textbf{SU} = Speedup compared to Always Retrieving (the higher, the better). Compared to the Always Retrieving baseline, \textsc{Repoformer}'s selective decisions improve both the accuracy and latency of these larger LMs.}
\vspace{-2mm}
\label{latency-transfer}
\end{table}

%% file: text/6_analysis.tex
\section{Analysis}
\label{section-analysis}

In this section, we present further analyses and ablation studies on \textsc{Repoformer-1B}.

\begin{figure}[t!]
\centering
\includegraphics[width=0.75\linewidth]{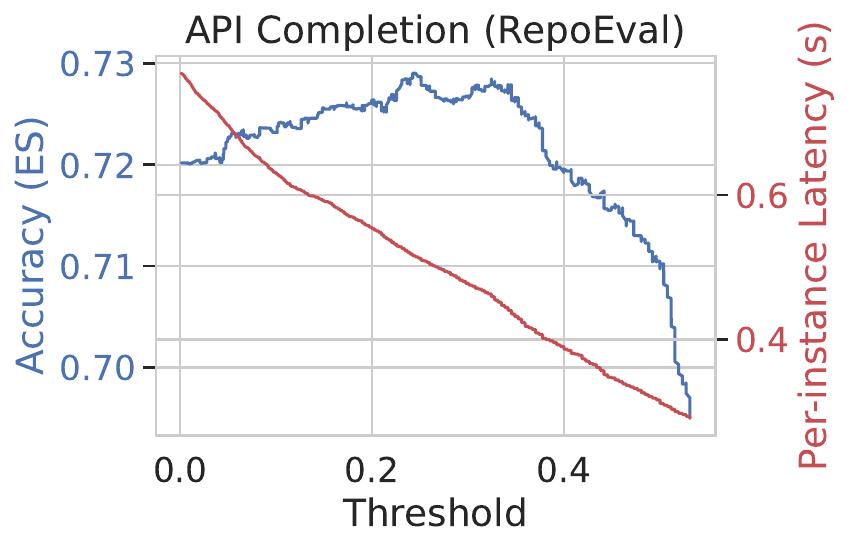}
\vspace{-2mm}
\caption{The accuracy and latency change with different threshold settings. Selective Retrieval with \textsc{Repoformer} achieves better accuracy and better latency than always retrieving. In addition, this behavior is relatively insensitive to the threshold. }
\label{threshold-sensitivity-api}
\end{figure}

\paragraph{Is \textsc{Repoformer} sensitive to threshold settings?} In \cref{threshold-sensitivity-api}, we present the code completion accuracy and latency of \textsc{Repoformer} as a function of the threshold. As the threshold increases, the model's code completion performance first increases due to avoiding potentially harmful retrievals. At threshold 0.4, the model still maintains similar performance compared to always retrieving, with latency reduced by 50\%. This result demonstrates that \textsc{Repoformer} can accommodate various threshold settings and provide a good accuracy-latency trade-off. We provide the visualization for other tasks in \cref{appendix-latency-accuracy}.

\begin{figure}[t!]
\centering
\includegraphics[width=\linewidth]{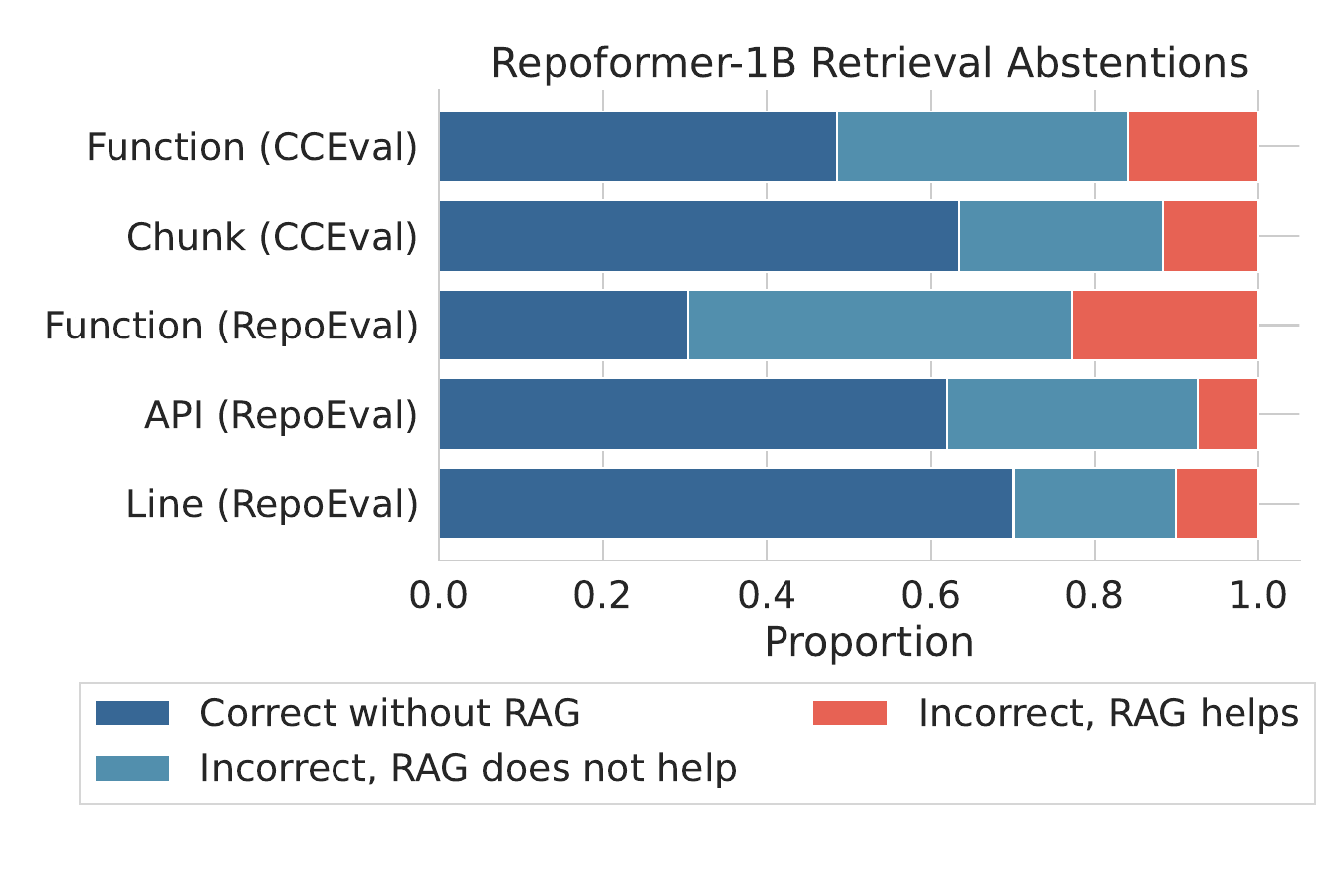}
\vspace{-8mm}
\caption{An analysis of the instances where \textsc{Repoformer-1B} abstains from retrieval. We divide the instances into (1) the model answering correctly without retrieval (dark blue), the model making a mistake that cannot be improved by retrieval (light blue), and the model achieving better performance when retrieval is performed (red). The precision of abstention is over 0.8 on all tasks except for Function (RepoEval), which has a precision of 0.78.}
\vspace{-2mm}
\label{abstention-analysis}
\end{figure}

\paragraph{Does \textsc{Repoformer} make accurate and calibrated selective retrieval decisions?} In \cref{abstention-analysis}, we evaluate the precision of retrieval abstention decisions made by \textsc{Repoformer}'s threshold selection strategy. We find that the abstentions are accurate for over 80\% instances across all the tasks: when \textsc{Repoformer} abstains from retrieval, its code completion prediction either is already correct without retrieval or cannot be improved by retrieval. We also evaluate the calibration of the selective decisions and find \textsc{Repoformer} generally making near-calibrated predictions for line and API completion while the calibration is suboptimal for function completion with UT employed as the metric (\cref{appendux-calibration}). We hypothesize that this could be caused by using ES to create the training signal and encourage future work to devise methods for labeling the quality of function completion more effectively.

\begin{figure}[t!]
\centering
\includegraphics[width=\linewidth]{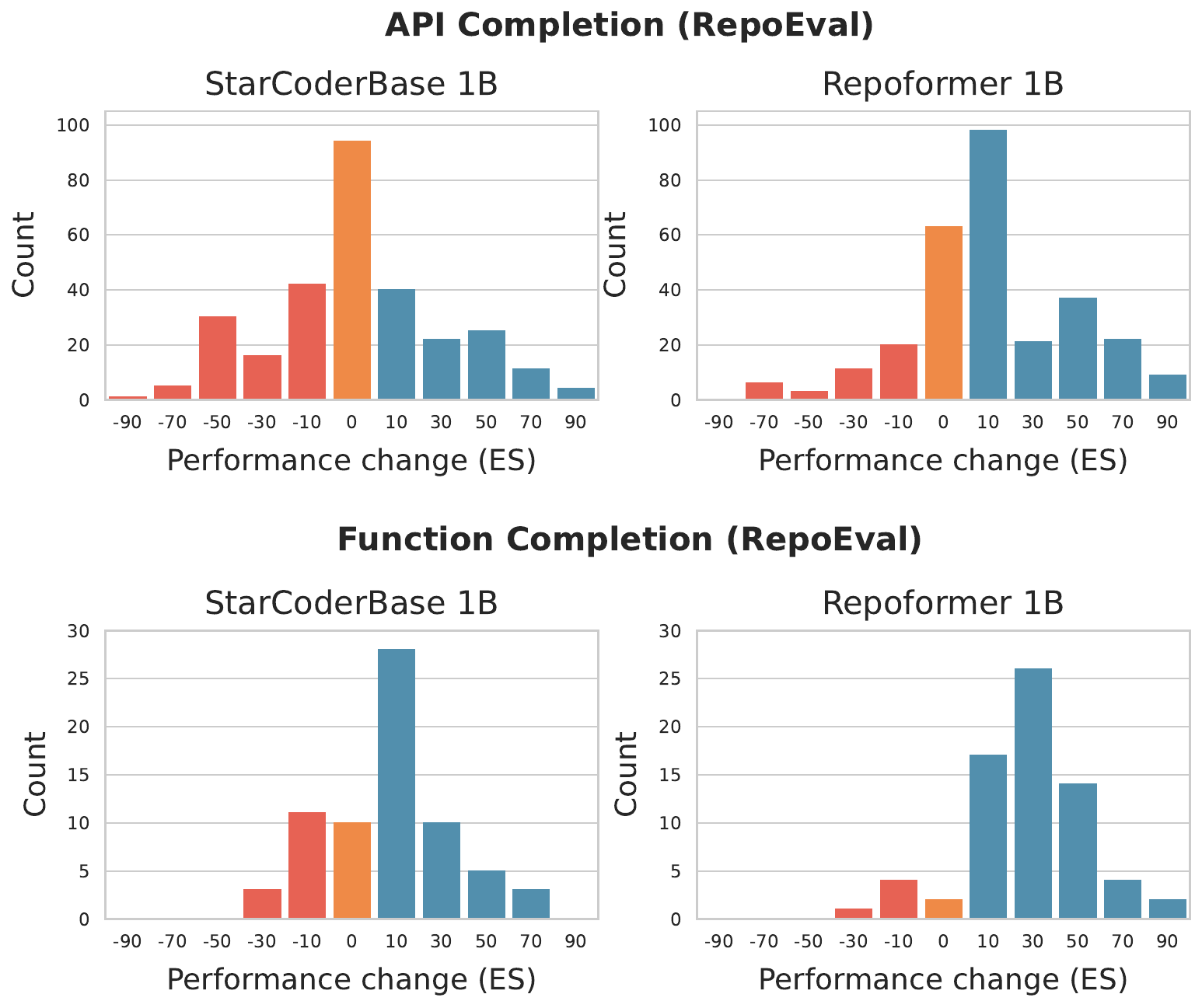}
\vspace{-3mm}
\caption{The performance change on RepoEval from retrieved cross-file context for the instances where \textsc{Repoformer} self-selects retrieval. Compared to StarCoderBase, \textsc{Repoformer} is better at leveraging $CC$ to improve the generation quality. }
\vspace{-2mm}
\label{leverage-rag}
\end{figure}

\paragraph{Is \textsc{Repoformer} robust to retrieval?} In Figure \ref{leverage-rag}, we show the performance change caused by $CC$ on the instances where \textsc{Repoformer} requests for retrieval. Compared to \textsc{StarCoderBase}, \textsc{Repoformer} exhibits more and greater performance gains upon observing $CC$. The number of performance decreases is also significantly reduced, indicating an improved robustness to the potentially irrelevant retrieval contexts. In \cref{retriever-robustness} in the appendix, we further study the effect of using \textit{dense retrieval}. Although dense retrieval returns an arguably different context distribution compared to sparse retrieval, \textsc{Repoformer} still exhibits strong improvements in both quality and latency.

\paragraph{Ablation Study} We study several alternative designs:
\begin{compactitem}
    \item (\textbf{A1}) Combining $\mathcal{L}_{eval}$ and $\mathcal{L}_{gen}$ as a single cross-entropy loss. In general, this down-weights $\mathcal{L}_{eval}$.
    \item (\textbf{A2}) Removing the self-evaluation loss $\mathcal{L}_{eval}$.
    \item (\textbf{A3}) Further removing all the $CC$ from A2. This amounts to only training on fill-in-the-middle.
    \item (\textbf{A4}) Placing \texttt{\textless{}cc\textgreater} and $CC$ after \texttt{\textless{}fim\_middle\textgreater} and marking its end with a new token \texttt{\textless{}cc\_end\textgreater}. A4 mainly studies whether it is more beneficial to train the LM to treat $CC$ as context fetched during fill-in-middle generation instead of part of the input context.
\end{compactitem}

We fine-tune StarCoderBase-1B with the same setup as \textsc{Repoformer} and present the results on CCEval in \cref{tab:ablation_study}. Although \textbf{A1} has slightly better RAG performance, it fails to make meaningful selective decisions due to $\mathcal{L}_{eval}$ being outweighed by $\mathcal{L}_{gen}$ in long sequences: the probability of \texttt{\textless cc\textgreater} is almost always 1. For \textbf{A2}, we find it only slightly outperforms \textsc{Repoformer}, suggesting learning $\mathcal{L}_{eval}$ does not harm the RAG ability a lot while bringing in the strong selective retrieval ability, which in turn boosts both accuracy and latency. \textbf{A3} has the same performance for in-file completion as \textsc{Repoformer}, but exhibits worse RAG performance, indicating the necessity of training with $CC$. Finally, \textbf{A4} achieves reasonable chunk completion performance but performs much worse in function completion. We hypothesize that placing $CC$ within the infilling part is  detrimental due to breaking the fill-in-the-middle semantics learned in StarCoder pre-training.


\input{tables/ablation}

%% file: tables/ablation.tex
\begin{table}[t!]
\centering
 \resizebox{0.99\linewidth}{!} {%
\begin{tabular}{c|c|ccc|ccc}
\hline
\multirow{2}{*}{\textbf{Model}} &\multirow{2}{*}{\textbf{RAG Policy}} & \multicolumn{3}{c|}{\textbf{Chunk Completion}}  & \multicolumn{3}{c}{\textbf{Function Completion}} \\
 & & \textbf{T} & \textbf{\%RAG} & \textbf{ES} & \textbf{T} & \textbf{\%RAG} & \textbf{ES} \\
 \hline
\multirow{2}{*}{\textbf{\textsc{SC}}} & No & - & 0\% & 60.09 & - & 0\% & 47.49 \\
& Always & - & 100\% & 63.73 & - & 100\% & 50.50 \\
\hline
\multirow{3}{*}{\textbf{\textsc{RF}}} & No & - & 0\% & 66.22 & - & 0\% & 49.77 \\
& Selective\textsubscript{T} & 0.20 & 75\% & 69.97 & 0.15 &	76\%&	53.71 \\
& Always & - & 100\% & 69.95 & - & 100\% & 53.56 \\
\hline
\multirow{3}{*}{\textbf{A1}}  & No & - & 0\% & 66.14 & - & 0\% & 49.25 \\
& Selective\textsubscript{T} & 0.99 & 100\% & 70.21 & 0.99 & 100\% & 53.93 \\
& Always & - & 100\% & 70.21 & - & 100\% & 53.93 \\
\hline
\multirow{2}{*}{\textbf{A2}} & No & - & 0\% & 66.49 & - & 0\% & 49.02 \\
& Always & - & 100\% & 70.45 & - & 100\% & 53.90 \\
\hline
\multirow{2}{*}{\textbf{A3}} & No & - & 0\% & 66.25 & - & 0\% & 49.01 \\
& Always & - & 100\% & 68.85 & - & 100\% & 52.12 \\
\hline
\multirow{3}{*}{\textbf{A4}} & No & - & 0\% & 64.96 & - & 0\% & 25.44 \\
& Selective\textsubscript{T} & 0.10 & 86\% & 69.35 & 0.10 & 83\% & 26.50 \\
 & Always & - & 100\% & 69.19 & - & 100\% & 26.35 \\
\hline
\end{tabular}
}
\caption{Ablation study results. We report the performance on two tasks from the CCEval dataset. \textbf{\textsc{SC}} = StarCoderBase-1B. \textbf{\textsc{RF}} = \textsc{Repoformer-1B}. \textbf{T} = threshold for the Selective\textsubscript{T} policy. We found T = 0.10 works better for A4 and thus applied it to all the A4 results. \textbf{\%RAG} = ratio of instances where RAG is performed.}
\vspace{-4mm}
\label{tab:ablation_study}
\end{table}

%% file: text/7_conclusion.tex
\section{Conclusion}

In this paper, we challenge the common assumption of always performing retrieval for RAG-based repository-level code completion. In response, we propose a selective retrieval augmentation framework powered by \textsc{Repoformer}, a code LM that identifies whether cross-file context is necessary, and self-triggers retrieval. Extensive evaluations demonstrate our approach's effectiveness in enhancing accuracy while significantly reducing latency, showcasing its potential in practical coding environments. 

\paragraph{Discussion} Building upon \textsc{Repoformer}, future research may consider several important directions:

\begin{compactenum}
    \item \textbf{Further speeding up large LMs.} Beyond as a selective retrieval policy, \textsc{Repoformer} has the potential to serve as an effective plug-in draft model in settings such as \textit{speculative decoding} \citep{chen2023accelerating}.
    \item \textbf{More effective function completion.} To enable a good scalability, we used lexical similarity as the signal for training label creation. Although this heuristics enables improvements in function completion evaluation, designing a more accurate and scalable labeling approach is an important future direction.
    \item \textbf{Personalized retrieval.} We apply a uniform selective policy across repositories. However, certain repositories could be inherently more RAG-friendly by exhibiting a higher level of duplication \citep{zhang2023repocoder}. Adapting the selective RAG paradigm towards accurate personalized policies is an important direction.
\end{compactenum}

\section*{Acknowledgement}
We express our gratitude to anonymous reviewers for their valuable suggestions to improve the quality of the paper. The authors also thank Amita Kamath and Po-Nien Kung for their constructive feedback provided during the paper writing process. Additionally, we would like to express gratitude to some other team members from Amazon CodeWhisperer and UCLANLP for their insightful discussions, which have contributed to the refinement of our work.

\section*{Impact Statement}
Our research introduces a novel approach to repository-level code completion that significantly enhances efficiency and accuracy by employing selective retrieval, reducing unnecessary computational waste, and contributing to more sustainable software development practices. Although promising in streamlining development workflows and potentially applicable in various domains, it is important to consider the implications of increased automation in software development, programming education, and the potential for inadvertent biases. Ensuring the ethical use and ongoing evaluation of such code automation technologies is crucial to maximize their societal benefits while mitigating risks. In addition, as a general infrastructure, it is important to design additional mechanisms that prevent RAG systems from revealing sensitive data in the retrieval database.

In this work, we mainly rely on open-sourced, permissively-licensed repositories (the Stack, CrossCodeEval) and models (StarCoder, CodeGen) to perform the experiments. However, as mentioned by \citet{ding2023crosscodeeval}, some of the repositories of RepoEval are with non-permissive licenses. We rely on the dataset and code distributed by the original RepoEval authors to perform the experiment and do not re-distribute the dataset or adapt it for other purposes.

%% file: text/appendix.tex
\newpage
\appendix
\onecolumn

\section{Detailed RAG Execution Setup}
\label{detailed-rag-setup}

Below, we describe the four steps we follow for executing RAG as well as the related hyperparameters.

\begin{enumerate}
    \item \textbf{Indexing.} All files in $F$ are divided into fix-sized code chunks with a sliding window. We set the chunk size to 20 for line, API, and chunk completion and set 50 for function completion. We use half of the chunk size as the stride size. Despite the duplication caused by the overlap between adjacent chunks, this design improves retrieval accuracy with tolerable cost, as the number of files is limited in a repository compared to large open-domain code corpora. 
    \item \textbf{Query Formation.} A query is constructed based on $X_l$. We always use a fixed number of lines at the end of $X_l$ (i.e., immediately preceding $Y$) as the query. The query contains the same number of lines as the chunks in the index.
    \item \textbf{Retrieval.} A similarity function $f$ is used to compare the query with every chunk and identify $k$ most similar code chunks. We use $k=10$ and Jaccard similarity \citep{jaccard1912distribution} for $f$ for the main results. Fragment alignment \citep{lu-etal-2022-reacc} is then applied: for each of the $k$ most similar code chunks, the chunk immediately following is included in $CC$ instead of the original chunk. We explored other choices mentioned in \cref{retriever_for_selective_retrieval} such as cosine similarity with UniXCoder \citep{guo-etal-2022-unixcoder} or CodeBLEU \citep{ren2020codebleu}, but find them failing to outperform Jaccard similarity.
    \item \textbf{Generation.} $CC$ is concatenated with the in-file context as a prompt for $\mathcal{M}$. The prompt is provided below.
\end{enumerate}

\paragraph{Prompt} Recent literature demonstrates the effectiveness of directly providing the retrieved information as part of the context of LMs \citep{ram2023ralm, shi2023replug}. Following these studies, we directly concatenate the in-file context with $CC$ to provide it to the model (\cref{main-framework}).  To prompt CodeGen-Mono, we use the following input ordering: 

\vspace{5pt}
\small
\centerline{\texttt{[Right Context] [Cross-file Context] [Left Context]}}
\normalsize

To prompt StarCoder, we use the following fill-in-the-middle-prompt:

\vspace{5pt}
\small
\centerline{\texttt{\textcolor{blue}{\textless fim\_prefix\textgreater} [Left Context] \textcolor{blue}{\textless fim\_suffix\textgreater} [Right Context] [Cross-file Context] \textcolor{blue}{\textless fim\_middle\textgreater}}}
\normalsize

For the cross-file contexts, we add a \texttt{\#} symbol to present them as comments and add the following line before each $cc_i$:

\vspace{5pt}
\centerline{\texttt{\# the below code fragment can be found in: [file path]}}

After concatenating the verbalized $cc_i$ together, we add another line to the start of the $CC$:

\vspace{5pt}
\centerline{\texttt{\# Here are some relevant code fragments from other files of the repo:}}

For the in-file completion baselines such as in \cref{80-20-rule} and \cref{section-why-infilling}, our prompts are exactly the previous prompts with the \texttt{[Cross-file Context]} part removed.

\paragraph{Decoding and Post-processing} For all the experiments, we follow previous work and use greedy search \citep{zhang2023repocoder, ding2023crosscodeeval}. We left-truncate the left context to 1024 tokens, right-truncate the right context to 512 tokens, and right-truncate the cross-file context to 512 tokens. The max generation length is set to 50 tokens for line, API, and chunk completion, and 256 tokens for function completion. We perform task-specific post-processing on the model's raw predictions. For line, API, and chunk completion, we truncate the prediction to having the same number of lines as in $Y$. For function completion, we first add a placeholder \texttt{pass} function body and use tree-sitter\footnote{\url{https://tree-sitter.github.io/tree-sitter/}} to determine the position of the function in the file. Then, we concatenate the $X_l$ and $\hat{Y}$, parse the string again with tree-sitter, and extract the function body as the final $\hat{Y}$ if the string can be parsed. Otherwise, we directly return the raw $\hat{Y}$ without post-processing.

\section{Why infilling?}
\label{section-why-infilling}

As part of the in-file context, $X_r$ contains rich information about how the future execution relies on the code to complete. Right contexts are also shown useful for tasks such as function call argument completion \citep{Pei_Zhao_Lausen_Zha_Karypis_2023}. However, previous literature such as \citet{zhang2023repocoder} suggests splitting $X_r$ and retrieving code chunks from it. With code LMs trained on fill-in-the-middle such as StarCoder, we argue that directly providing $X_r$ in the prompt is more preferable.  

\begin{figure}[h]
\centering
\includegraphics[width=\linewidth]{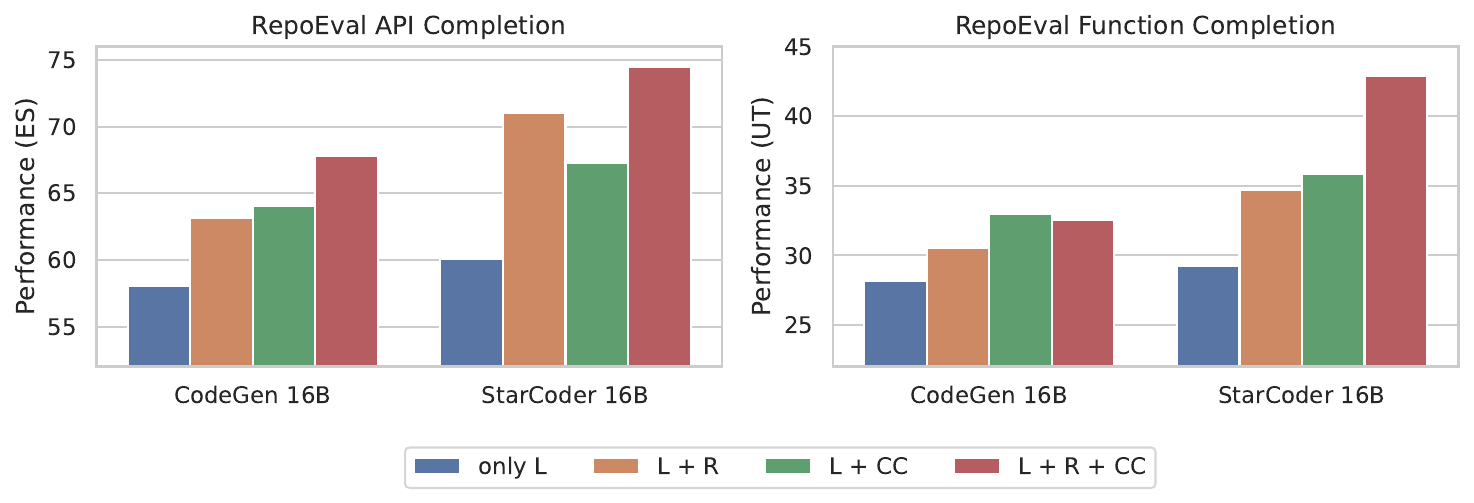}
\caption{A comparison between four prompting strategies for RepoEval by combining left context (L), right context (R), and cross-file contexts (CC). Leveraging right contexts to build infilling-style prompt generally improves the performance regardless whether CC is present or not. StarCoder exhibits larger gains from right contexts, potentially due to its fill-in-the-middle pre-training.}
\label{completion_vs_infilling}
\end{figure}

To illustrate, we investigate the effect of directly providing $X_r$ in the prompt for CodeGen-Mono 16B and StarCoder on current-file code completion and retrieval-augmented code completion. Figure \ref{completion_vs_infilling} presents the performance on RepoEval with different types of contexts provided in the prompt. Whether cross-file contexts are present or not, providing right contexts can greatly improve the code completion performance. The gain is consistent for both API and function completion. Compared to CodeGen, StarCoder can better leverage the right context to generate more accurate code. Overall, we observe that leveraging the entire right context to perform infilling represents a much stronger baseline. Therefore, in this paper we have exclusively focused on the infilling setting with the StarCoder family.

\section{Trial Retrieval and Trial Generation}
\label{appendix-retriever-for-srag}

In this section, we present a detailed evaluation of two selective RAG strategies: trial retrieval and trial generation.

\subsection{Trial Retrieval} 
To gauge the relevance of retrieved context, using the similarity scores from the retrievers is a natural option. In this section, we investigate \textit{trial retrieval} as a baseline for informing the decisions for selective RAG. We apply three off-the-shelf retrievers on RepoEval. For each retriever, we score each of the instances with the similarity between the top-1 retrieved code chunk and the query. The score is compared to a threshold decide whether the prompt should feature $CC$ or not. If score is higher than the threshold, we use top-10 code chunks retrieved by the same retriever as the cross-file context. We consider the following three retrievers:

\begin{itemize}
    \item \textbf{jaccard}: the Jaccard index \citep{jaccard1912distribution}.
    \item \textbf{weighted\_ngram}: the weighted n-gram matching term introduced in the CodeBLEU metric \citep{ren2020codebleu}. 
    \item \textbf{unixcoder}: the cosine similarity of UniXcoder embedding \citep{guo-etal-2022-unixcoder}. 
\end{itemize}

\begin{figure*}[t!]
\centering
\includegraphics[width=\textwidth]{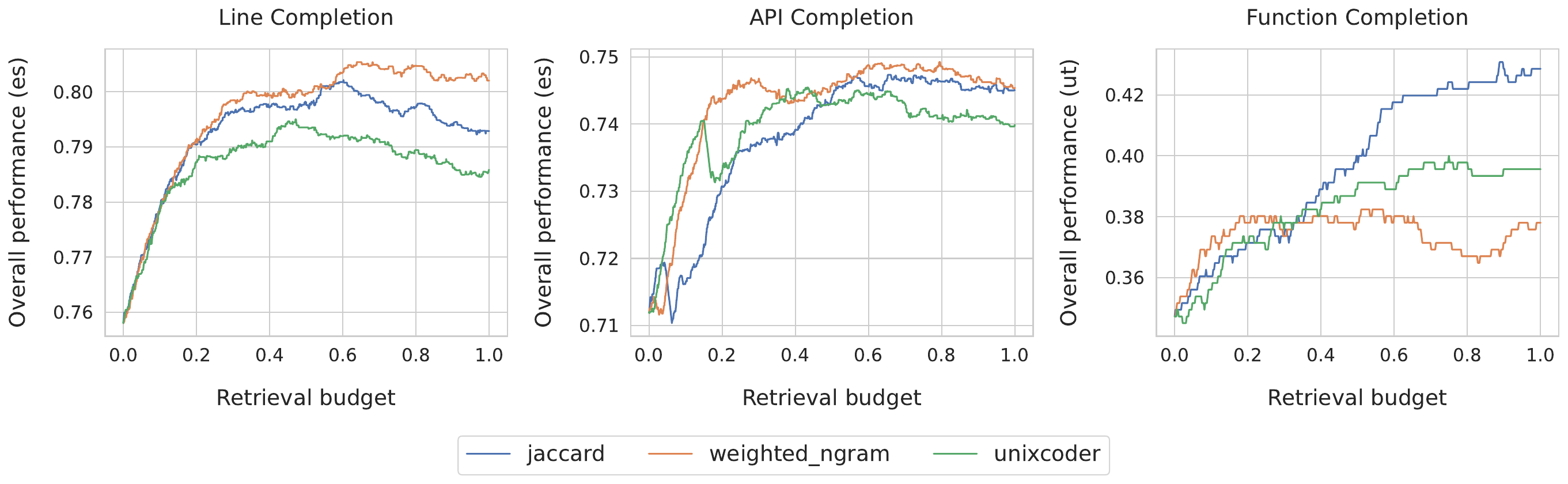}
\caption{A comparison of the effectiveness of different similarity functions for selective RAG with StarCoder 16B. We plot the retrieval budget in the x-axis, which is the percentage of instances to perform retrieval. We report score on the entire testing dataset for each budget. Specifically, the retriever's similarity score is used select a subset to perform retrieval, and for the other instances in-file completion is performed without retrieval. In most of the cases, 40\% retrieval can be saved without sacrificing the code completion performance.  }
\vspace{-2mm}
\label{retriever_for_selective_retrieval}
\end{figure*}

Figure \ref{retriever_for_selective_retrieval} presents the selective RAG performance of StarCoder under different budgets.We observe that the retrievers' similarity scores serve as a promising signal for deciding whether the retrieved information can improve the RAG performance.  For most retrievers and tasks, the performance of full retrieval could be reached with at most 60\% retrieval budget. This trend also aligns with the remark in \citet{zhang2023repocoder} on the correlation between in-repository duplication and the gain from $CC$. However, it is worth noting that this strategy brings no latency gain as it still implements always retrieving. In addition, the knowledge of whether the LM could be benefited by the retrieved context is not leveraged. 

\begin{figure*}[t!]
\centering
\includegraphics[width=\textwidth]{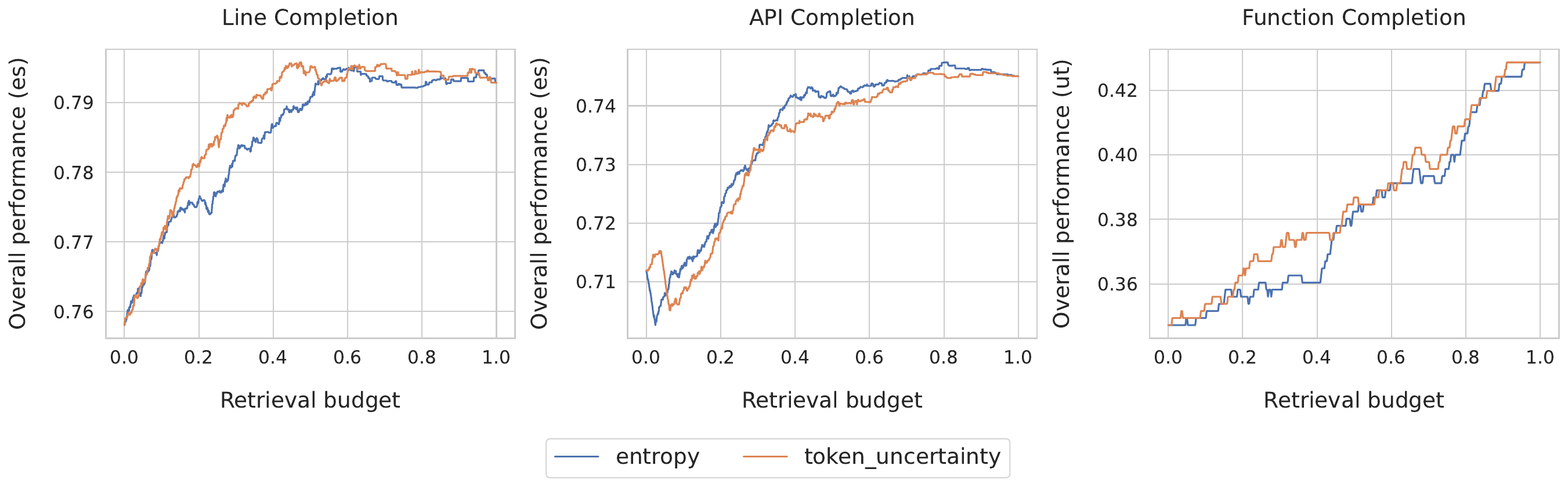}
\caption{A comparison of the effectiveness of two uncertainty metrics for selective RAG with StarCoder 16B. We plot the retrieval budget in the x-axis and report score on the entire testing dataset for each budget. We observe that the uncertainty-based metrics fail for long sequence generation such as function completion. Token uncertainty outperforms entropy for line completion while entropy is slightly better for API completion. Overall, we find that uncertainty-based selective RAG is not as effective as retriever-based (\cref{retriever_for_selective_retrieval}).}
\vspace{-2mm}
\label{uncertainty_for_selective_retrieval}
\end{figure*}

\subsection{Trial Generation} 

Next, we evaluate two uncertainty-based selective RAG strategies that have been explored by previous works.

\begin{itemize}
    \item \textbf{entropy}: the sequence-level entropy as used in \citet{li-etal-2023-web}. We estimate the entropy by performing vanilla sampling for 20 times without any temperature scaling or distribution truncation.
    \item \textbf{token uncertainty}: the probability of the most unlikely token in the sequence decoded with greedy search, as used in \citet{jiang2023active}. This metric can be seen as the lower bound of the per-token maximum probability.
\end{itemize}

Figure \ref{uncertainty_for_selective_retrieval} presents the selective RAG performance of StarCoder under different budgets, similar to the previous evaluation setting. We find that the selective RAG performance of uncertainty-based metrics is inconsistent across sequence lengths. As the length of $\hat{Y}$ increases (from line to API, and form API to function), the effectiveness of uncertainty-based metrics drops significantly. In addition, the selective performance cannot outperform the methods based on trial retrieval.

\section{Data Creation for \textsc{Repoformer} Training and CrossCodeLongEval}
\label{appendix-data-creation-algo}
\label{appendix-training-data-analysis}
\label{appendix-cceval-creation}

We present the full self-supervised data creation algorithm in \cref{alg:data-creation-chunk} (for chunk completion data) and \cref{alg:data-creation-function} (for function completion data). $R_{filtered}$ stands for the remaining repositories after applying the filtering criteria in \cref{section-repoformer-method}. In the next section, we present further analyses on the training data distribution.

\input{figures/data_creation_algos}

\begin{figure*}[h]
\centering
\includegraphics[width=\textwidth]{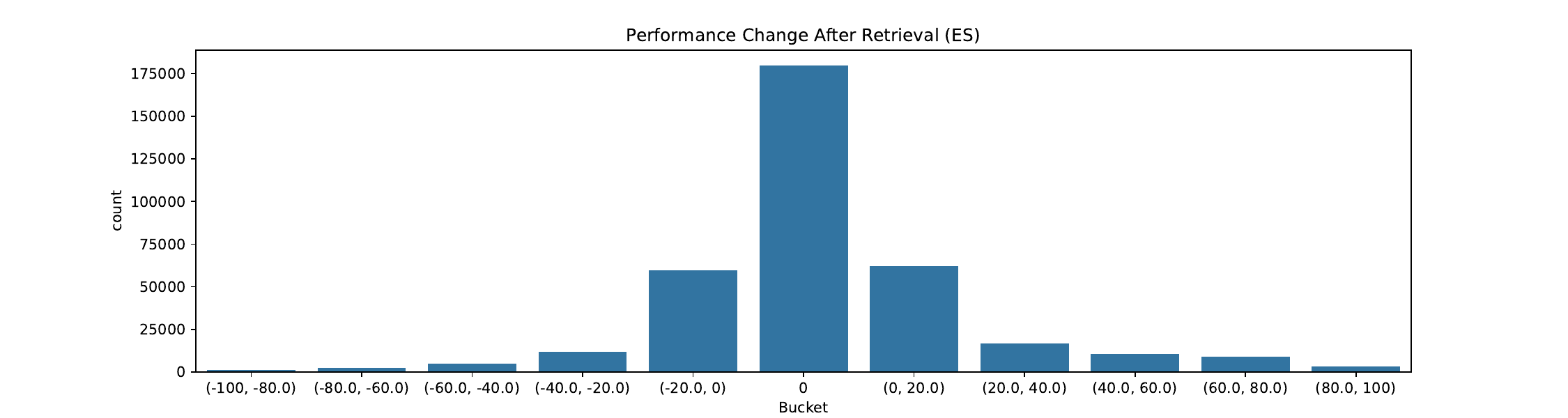}
\caption{The performance gain on \textsc{Repoformer} training data exhibited by StarCoderBase-1B from retrieved cross-file context. The sign of the performance change is used to generate the label for \textsc{Repoformer} training. Each (start, end) bucket contains values ranging from start to end except for the central bucket, which corresponds to exactly 0.}
\label{repoformer_training_data_es_change}
\end{figure*}

\paragraph{Training Data Analysis} For the 240k chunk completion and 120k function completion instances, we plot the performance change after providing $CC$ in \cref{repoformer_training_data_es_change}. In total, 30.18\% chunk completion instances and 35.16\% function completion instances are labeled with positive (i.e., retrieval should be triggered). The average length of $Y$ is 3.53 lines for chunk completion and 11.77 lines for function completion. 

\paragraph{CrossCodeLongEval Construction} One drawback of RepoEval is its limited repository coverage. To verify the performance on diverse repositories, we collect and curate a new evaluation dataset for repository-level code completion. 

\begin{itemize}
    \item \textbf{Repository collection.} We first solicited 1744 raw Python repositories from the authors of CrossCodeEval \citep{ding2023crosscodeeval}. These repositories were created between 2023-03-05 to 2023-06-15 and collected on 2023-09-01. They have been ensured to not overlap with the Stack \citep{kocetkov2022stack}. 
    \item \textbf{Target line sampling.} We avoided using the CrossCodeEval benchmark as the original benchmark explicit removed the instances where StarCoderBase-1B can correctly answer without the retrieved context. To simulate a more natural distribution of code completion, we sample new blanks from the raw repositories. Specifically, we run \cref{alg:data-creation-chunk} and \cref{alg:data-creation-function} to gather chunk completion and function completion instances.
    \item \textbf{Data analysis} In \cref{tab:dataset_stats}, we present the basic statistics of RepoEval and CrossCodeLongEval.
\end{itemize}

\input{tables/dataset_stats}

\newpage
\section{Extended Analyses}

\subsection{Calibration of \textsc{Repoformer}'s Selective Retrieval Prediction}
\label{appendux-calibration}

\begin{figure}[h!]
\centering
\includegraphics[width=\linewidth]{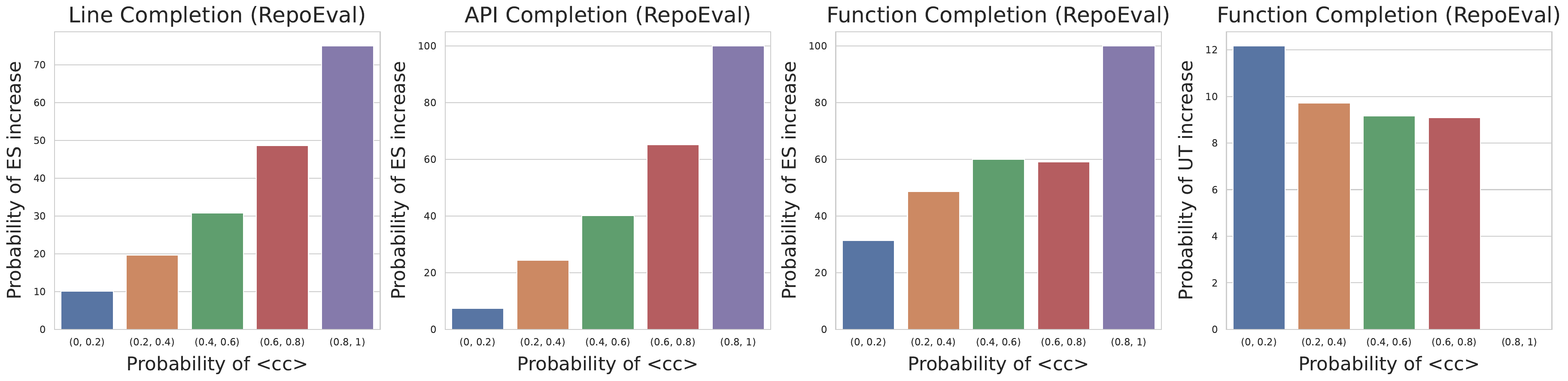}
\caption{The calibration of selective retrieval predictions. \textsc{Repoformer} makes generally calibrated predictions when ES is used as the metric and the generation is of moderate lengths. The prediction is not calibrated for function completion when the metric is UT.}
\vspace{2mm}
\label{calibration}
\end{figure}

We evaluate the calibration of \textsc{Repoformer-1B}'s selective decisions. \cref{calibration} plots the probability of \texttt{\textless{}cc\textgreater} against the probability of the model's performance could be improved by the $CC$, measured by comparing the prediction with and without $CC$. When ES is used as the evaluation metric, \textsc{Repoformer-1B} generally makes near-calibrated predictions for Line and API Completion. However, when it comes to longer-formed function completion, especially when UT is employed as the metric, \textsc{Repoformer-1B}'s predictions are not calibrated. One possible reason is the use of ES as the training signal. We encourage future work to devise methods for effectively labeling the correctness of function completion. In addition, future work should consider training \textsc{Repoformer} to perform multiple self-assessments for long-form generations.

\subsection{CrossCodeEval and Multilingual \textsc{Repoformer}}
\label{appendux-crosscodeeval-results}
\label{appendix-multilingual-results}
\label{appendix-7b-results}

This section provides additional results on the 4-language original CrossCodeEval test set \citep{ding2023crosscodeeval}. We choose to not present the results in the main text as the data creation process of CrossCodeEval explicitly selected the instances where cross-file information is generally required, thus making the contributions from selective retrieval incomplete. On this dataset, we evaluate StarCoder, \textsc{Repoformer-1B/3B/7B} trained on Python and \textsc{Repoformer-M} trained on multilingual repository-level code completion. Despite the setup difference, we are still able to observe substantial performance gains.


\paragraph{Multilingual \textsc{Repoformer}} We experimented with applying the \textsc{Repoformer} training scheme to multiple languages. Specifically, we collect public Python, Java, C\#, and TypeScript repositories from the Stack \citep{kocetkov2022stack} that contain at least 20 files and 20,000 lines of code. We do not apply the local import criteria due to implementation difficulties. Then, we follow the algorithm described in \cref{appendix-data-creation-algo} to create 90k chunk completion and 30k function completion instances per language. Using this dataset, we fine-tune StarCoderBase following the setup described in \cref{repoformer-implementation-details} (same infrastructure and hyperparameters). We call this model \textsc{Repoformer-M}.

\paragraph{Evaluation Results} We present the results on CrossCodeEval in \cref{crosscodeeval} and summarize the observations below:

\begin{itemize}
    \item \textbf{Strong cross-lingual transfer.} \textsc{Repoformer} trained on Python data achieves strong performance across multiple languages, including three languages it is not fine-tuned on. The result highlights the generalizability of the learned self-evaluation and robust code completion abilities. 
    \item \textbf{Multi-lingual \textsc{Repoformer}.}  \textsc{Repoformer-M} outperforms the same-sized \textsc{StarCoderBase} by a large margin. For the 1B, 7B, \textsc{Repoformer-M} outperforms \textsc{Repoformer} by a small margin. For 3B, the two models give similar performance. This is reasonable as the two models are learned on similar sized training data.
\end{itemize}

\input{tables/crosscodeeval}

\newpage
\subsection{\textsc{Repoformer}'s Robustness to the Retriever Choice}
\label{appendix-retriever-robustness}

In this section, we investigate the performance of \textsc{Repoformer} with the cosine similarity of UniXcoder embedding \citep{guo-etal-2022-unixcoder} as the retriever instead of Jaccard similarity. As shown in \cref{retriever-robustness}, we are able to observe similar patterns compared to \cref{latency-repoformer}: selective retrieval is able to improve both the accuracy and the latency of the entire RAG system. In addition, as retrieval consumes a larger proportion of latency than when sparse retriever is used, selective retrieval brings more substantial performance gains, with threshold selection bringing more than 70\% speedup. 

\input{tables/retriever_robustness}

\subsection{Full Latency-Accuracy Visualizations}
\label{appendix-latency-accuracy}

\begin{figure}[t!]
\centering
\includegraphics[width=\textwidth]{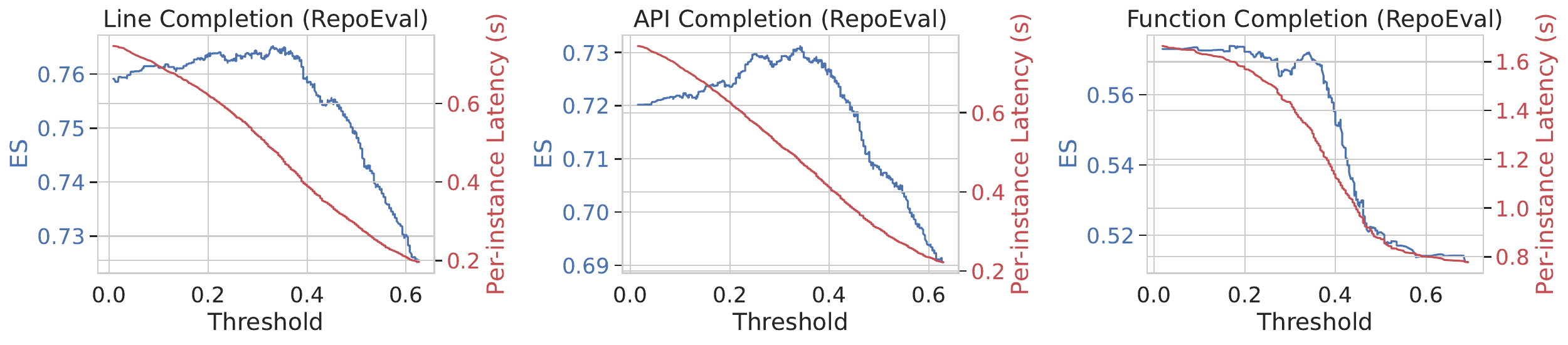}
\caption{Latency-accuracy trade-off of self-selective RAG for \textsc{Repoformer-1B}.}
\label{full-latency-accuracy-tradeoff-repoformer-1b}
\end{figure}

\begin{figure}[t!]
\centering
\includegraphics[width=\textwidth]{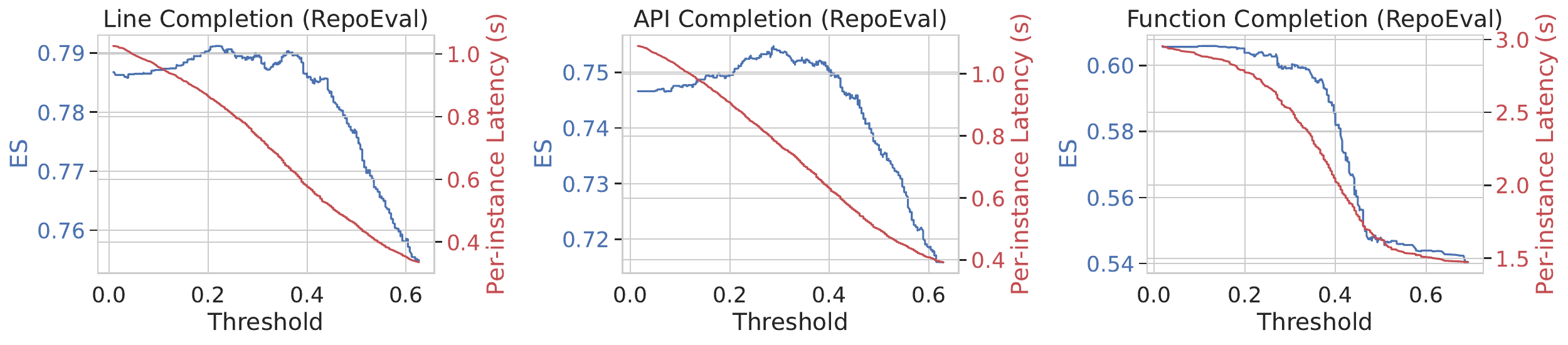}
\caption{Latency-accuracy trade-off of self-selective RAG for \textsc{Repoformer-3B}.}
\label{full-latency-accuracy-tradeoff-repoformer-3b}
\end{figure}

\begin{figure}[t!]
\centering
\includegraphics[width=\textwidth]{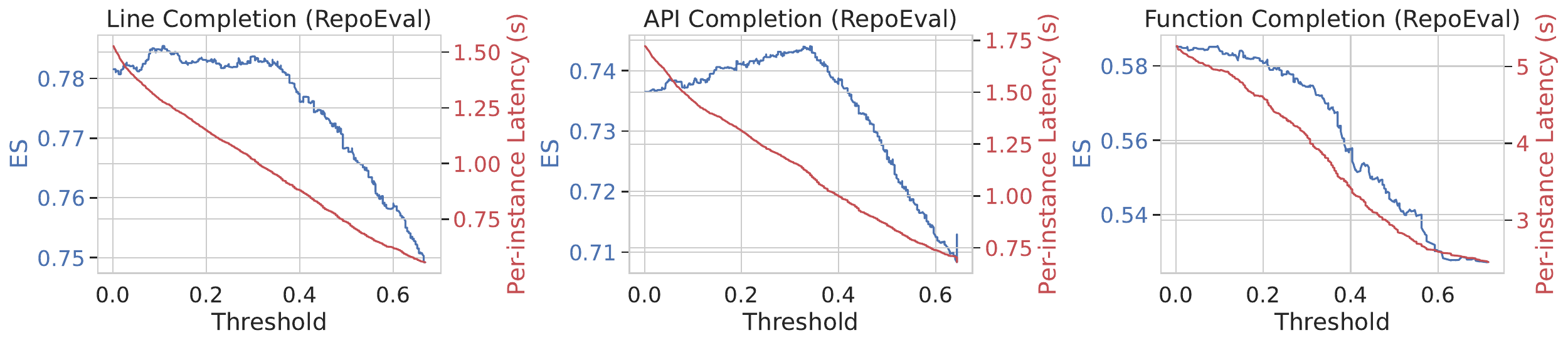}
\caption{Latency-accuracy trade-off of selective RAG for \textsc{StarCoderBase-7B}. \textsc{Repoformer-1B} is used for the selective decisions.}
\label{full-latency-accuracy-tradeoff-starcoder-7b}
\end{figure}

\begin{figure}[t!]
\centering
\includegraphics[width=\textwidth]{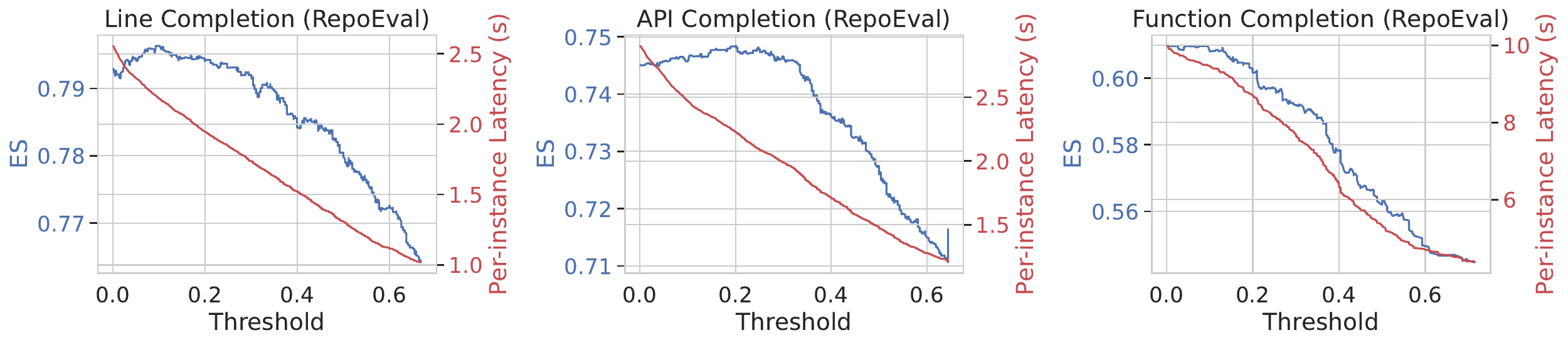}
\caption{Latency-accuracy trade-off of selective RAG for \textsc{StarCoder}. \textsc{Repoformer-1B} is used for the selective decisions.}
\label{full-latency-accuracy-tradeoff-starcoder-16b}
\end{figure}

In this section, we present the latency-accuracy trade-off plots for \textsc{Repoformer-1B}, \textsc{Repoformer-3B}, \textsc{StarCoderBase-7B}, and \textsc{StarCoder} on the three tasks from RepoEval. We use self-selective RAG for the \textsc{Repoformer} models and for \textsc{StarCoder}, we use \textsc{Repoformer-1B} to make the selective RAG decisions. The results are presented in \cref{full-latency-accuracy-tradeoff-repoformer-1b} to \cref{full-latency-accuracy-tradeoff-starcoder-16b}. Overall, we observe that no matter for self-selective RAG or making selective predictions for a larger model, \textsc{Repoformer} is able to improve the accuracy and latency at the same time. The improvement is more apparent in the line and API completion tasks. For function completion, as discussed in the main text, RepoEval uses very small repositories to enable easy unit testing. As a result, the retrieval overhead is low in general and thus does not significantly affect the latency of the entire RAG system.

\newpage

%% file: figures/data_creation_algos.tex
\begin{algorithm}
\caption{\textsc{Repoformer} Training  Data Creation (Chunk Completion)}
\label{alg:data-creation-chunk}
\begin{algorithmic}
\STATE {\bfseries Input:} Filtered set of repositories \( R_{filtered} \), language model \( \mathcal{M} \), label threshold \( T \)
\STATE {\bfseries Output:} chunk completion training dataset $\mathcal{D}$
\STATE $\mathcal{D} \gets \emptyset$
\FOR{each $r \in R_{filtered}$} 
    \STATE $\mathcal{D}_r \gets \emptyset$
    \STATE $\mathcal{C}_{raw} \gets$ Break $r$ into non-overlapping chunks of 10 lines each
    \STATE $\mathcal{C}_r \gets$ Cluster $\mathcal{C}_{raw}$ with KMeans using TF-IDF features, with the constraint $|\mathcal{C}_r| = 0.2|\mathcal{C}_{raw}|$
    \FOR{each $c \in \mathcal{C}_r$} 
        \STATE $k \sim \text{Poisson}(\lambda = 3)$
        \STATE $s \gets$ Randomly sample a chunk from $c$
        \STATE $Y \gets$ Randomly cut a sub-chunk from $s$ that spans $k$ consecutive lines
        \STATE $ X_l, X_r \gets$ Recover the in-file left context and right context corresponding to $Y$
        \IF{$rand(0, 1) > 0.5$}
            \STATE $\mathcal{Q} \gets$ Concatenate(last $5k$ lines of $X_l$, $Y$, first $5k$ lines of $X_r$) \hfil // query formation
        \ELSE
            \STATE $\mathcal{Q} \gets$ Concatenate(last $5k$ lines of $X_l$, first $5k$ lines of $X_r$)
        \ENDIF
        
        \STATE $CC \gets$ Retrieve top-3 cross-file contexts from \( r \) using \( \mathcal{Q} \) via jaccard similarity, each of length $10k$   
         
        \STATE $\hat{Y}_{base} \gets \mathcal{M}(X_l, X_r)$
        \STATE $\hat{Y}_{RAG} \gets \mathcal{M}(X_l, X_r, CC)$
        \STATE $label \gets ES(\hat{Y}_{RAG}, Y) - ES(\hat{Y}_{base}, Y) > T$ \hfil // boolean value
        \STATE Append $(X_l, X_r,\ Y,\ CC,\ label)$ to $\mathcal{D}_r$
    \ENDFOR
    \STATE $\mathcal{D} \gets \mathcal{D} \cup \mathcal{D}_r$
\ENDFOR
\end{algorithmic}
\end{algorithm}

\begin{algorithm}
\caption{\textsc{Repoformer} Training Data Creation (Function Completion)}
\label{alg:data-creation-function}
\begin{algorithmic}
\STATE {\bfseries Input:} Filtered set of repositories \( R_{filtered} \), language model \( \mathcal{M} \), label threshold \( T \)
\STATE {\bfseries Output:} function completion training dataset $\mathcal{D}$
\STATE $\mathcal{D} \gets \emptyset$
\FOR{each $r \in R_{filtered}$}
    \STATE $\mathcal{D}_r \gets \emptyset$
    \STATE $\mathcal{C}_{raw} \gets$ Gather all the functions between 3 and 30 lines
    \STATE $\mathcal{C}_r \gets$ Cluster $\mathcal{C}_{raw}$ with KMeans using TF-IDF features, with the constraint $|\mathcal{C}_r| = 0.2|\mathcal{C}_{raw}|$
    \FOR{each $c \in \mathcal{C}_r$} 
        \STATE $s \gets$ Randomly sample a function from $c$
        \STATE $Y \gets$ Cut only the body part of the function
        \STATE $ X_l, X_r \gets$ Recover the in-file left context and right context corresponding to $Y$
        \IF{$rand(0, 1) > 0.5$}
            \STATE $\mathcal{Q} \gets$ Concatenate(last $20$ lines of $X_l$, $Y$, first $20$ lines of $X_r$)
        \ELSE
            \STATE $\mathcal{Q} \gets$ Concatenate(last $20$ lines of $X_l$, first $20$ lines of $X_r$)
        \ENDIF
        
        \STATE $CC \gets$ Retrieve top-3 cross-file contexts from \( r \) using \( \mathcal{Q} \) via jaccard similarity, each of length $10k$

        \STATE $\hat{Y}_{base} \gets \mathcal{M}(X_l, X_r)$
        \STATE $\hat{Y}_{RAG} \gets \mathcal{M}(X_l, X_r, CC)$
        \STATE $label \gets ES(\hat{Y}_{RAG}, Y) - ES(\hat{Y}_{base}, Y) > T$ \hfil // boolean value
        \STATE Append $(X_l, X_r,\ Y,\ CC,\ label)$ to $\mathcal{D}_r$
    \ENDFOR
    \STATE $\mathcal{D} \gets \mathcal{D} \cup \mathcal{D}_r$
\ENDFOR
\end{algorithmic}
\end{algorithm}

%% file: tables/dataset_stats.tex
\begin{table}[h]
\centering
 \resizebox{0.6\linewidth}{!} {%
\begin{tabular}{c|ccc|cc}
\hline
 & \multicolumn{3}{c|}{\textbf{RepoEval}}  & \multicolumn{2}{c}{\textbf{CrossCodeLongEval}} \\
 & \textbf{Line} & \textbf{API} & \textbf{Function} & \textbf{Chunk} & \textbf{Function} \\
\hline
\# repositories & 16 & 16 & 16 & 944 & 1460 \\
\# instances & 1600 & 1600 & 455 & 5000 & 5000 \\
$|X_l|_{line}$ & 30.7 & 30.8 & 31.1 & 24.7 & 31.7 \\
$|X_l|_{token}$ & 796.3 & 890.7 & 761.1 & 661.9 & 672.1 \\
$|X_r|_{line}$ & 15.1 & 13.9 & 16.2 & 12.9 & 14.4 \\
$|X_r|_{token}$ & 449.9 & 430.4 & 412.4 & 404.2 & 371.3 \\
$|Y|_{line}$ & 1.0 & 2.1 & 7.8 & 1.47 & 9.5 \\
$|Y|_{token}$ & 12.0 & 25.4 & 97.8 & 19.2 & 111.2\\
\hline
\end{tabular}
}
\caption{Descriptive statistics of RepoEval and CrossCodeLongEval. For $|Y|$, $|X_l|$, and $|X_r|$, we report both the number of lines as well as the number of tokens (using the StarCoder tokenizer) in the groundtruth, left context, and the right context.}
\label{tab:dataset_stats}
\end{table}

%% file: tables/crosscodeeval.tex
\begin{table}[t!]
\centering
\resizebox{\linewidth}{!} {
\begin{tabular}{c|c|cc|cc|cc|cc}
\hline
\multirow{2}{*}{\textbf{Model}} & \multirow{2}{*}{\textbf{RAG Policy}} & \multicolumn{2}{c|}{\textbf{Python}} & \multicolumn{2}{c|}{\textbf{Java}} & \multicolumn{2}{c|}{\textbf{C\#}} & \multicolumn{2}{c}{\textbf{TypeScript}} \\
 & & \textbf{Code ES} & \textbf{ID F1} & \textbf{Code ES} & \textbf{ID F1} & \textbf{Code ES} & \textbf{ID F1} & \textbf{Code ES} & \textbf{ID F1} \\
 \hline
\multirow{2}{*}{\textsc{StarCoderBase-1B}} & No & 68.83 & 58.18 & 73.60 & 63.69 & 79.30 & 66.40 & 67.09 & 60.15 \\
 & Always & 71.57 & 62.42 & 74.54 & 65.83 & 79.04 & 66.82 & 67.66 & 60.60 \\
 \hdashline
 \textsc{Repoformer-1B} & Selective\textsubscript{T} & 71.29 & 62.81 & 75.12 & 67.16 & 83.08 & 74.24 & 69.90 & 64.07 \\
  \textsc{Repoformer-M-1B} & Selective\textsubscript{T} & 71.55 & 62.89 & 75.92 & 67.86 & 84.44 & 76.00 & 70.07 & 64.41 \\
  \hline
\multirow{2}{*}{\textsc{StarCoderBase-3B}} & No & 71.07 & 61.63 & 76.10 & 67.56 & 81.46 & 69.95 & 70.56 & 64.83 \\
& Always & 73.65 & 65.93 & 77.52 & 70.15 & 81.75 & 71.26 & 70.91 & 65.09 \\
 \hdashline
\textsc{Repoformer-3B} & Selective\textsubscript{T} & 74.57 & 66.86 & 78.40 & 71.26 & 85.92 & 78.62 & 73.70 & 68.66 \\
 \textsc{Repoformer-M-3B} & Selective\textsubscript{T} & 73.80 & 66.72 & 77.68 & 71.01 & 85.31 & 77.70 & 72.51 & 67.06\\
 \hline
 \multirow{2}{*}{\textsc{StarCoderBase-7B}} & No & 72.47 & 63.76 & 77.21 & 68.97 & 83.06 & 72.06 & 72.34 & 67.06 \\
 & Always & 75.02 & 67.69 & 77.70 & 70.57 & 83.64 & 74.39 & 73.01 & 67.56 \\
 \hdashline
\textsc{Repoformer-7B} &  Selective\textsubscript{T} & 75.34 & \textbf{68.27} & 78.90 & 72.35 & 83.80 & 76.88 & 73.59 & 69.10 \\
  \textsc{Repoformer-M-7B} & Selective\textsubscript{T} & \textbf{75.35} & 67.88 & \textbf{79.11} & \textbf{72.82} & \textbf{86.53} & \textbf{79.77} & \textbf{74.60} & \textbf{70.01} \\
\hline
\end{tabular}
}
\caption{Evaluation results on CrossCodeEval. We report edit similarity for code matching as well as the F1 score for identifier matching. The best scores across all models are boldfaced. }
\label{crosscodeeval}
\end{table}

%% file: tables/retriever_robustness.tex
\begin{table}[h!]
 \centering
 \resizebox{0.8\linewidth}{!} {
\begin{tabular}{c|c|ccc|ccc}
\hline
\multirow{2}{*}{\textbf{Model}} & \multirow{2}{*}{\textbf{RAG Policy}} & \multicolumn{3}{c|}{\textbf{API Completion}} & \multicolumn{3}{c}{\textbf{Line Completion}}  \\
& & \textbf{ES} & \textbf{\%RAG} & \textbf{SU} & \textbf{ES} & \textbf{\%RAG} & \textbf{SU} \\
\hline
& Always & 71.69 & $100\%$ & $0\%$ & 75.25 & $100\%$ & 0\% \\
& Selective\textsubscript{G}  & 70.82 & $18\%$ & $71\%$ & 73.70 & $19\%$  & $71\%$ \\
\multirow{-3}{*}{\textsc{Repoformer-1B}} & \multicolumn{1}{c|}{Selective\textsubscript{T}} & 72.39 & $61\%$ & $33\%$ & 75.65 & $62\%$ & $33\%$ \\
\hline
& Always  & 74.48 & $100\%$ & $0\%$ & 78.24 & $100\%$ & 0\% \\
& Selective\textsubscript{G}  & 73.26 & $19\%$  & $65\%$ & 76.74 & $20\%$ & $66\%$ \\
\multirow{-3}{*}{\textsc{Repoformer-3B}} & \multicolumn{1}{c|}{Selective\textsubscript{T}} & 74.69 & $78\%$ & $21\%$ & 78.63  & $74\%$ & $31\%$  \\
\hline
\end{tabular}
}
\caption{RAG performance of \textsc{Repoformer} with two self-selective RAG paradigms and dense retrieval used instead of Jaccard similarity. \textbf{\%RAG} = ratio of instances where RAG is performed. \textbf{SU} = Speedup compared to always retrieving. Compared to the always retrieving baseline, the Selective\textsubscript{T} strategy consistently demonstrates gains in both accuracy and latency. The Selective\textsubscript{G} strategy shows much larger latency gains with a small performance degradation. Compared to sparse retrieval, we observe more substantial latency gains.}
\label{retriever-robustness}
\end{table}

%% file: main.bbl
\begin{thebibliography}{41}
\providecommand{\natexlab}[1]{#1}
\providecommand{\url}[1]{\texttt{#1}}
\expandafter\ifx\csname urlstyle\endcsname\relax
  \providecommand{\doi}[1]{doi: #1}\else
  \providecommand{\doi}{doi: \begingroup \urlstyle{rm}\Url}\fi

\bibitem[Asai et~al.(2024)Asai, Wu, Wang, Sil, and Hajishirzi]{asai2023self}
Asai, A., Wu, Z., Wang, Y., Sil, A., and Hajishirzi, H.
\newblock Self-{RAG}: Learning to retrieve, generate, and critique through self-reflection.
\newblock In \emph{The Twelfth International Conference on Learning Representations}, 2024.
\newblock URL \url{https://openreview.net/forum?id=hSyW5go0v8}.

\bibitem[Bavarian et~al.(2022)Bavarian, Jun, Tezak, Schulman, McLeavey, Tworek, and Chen]{bavarian2022efficient}
Bavarian, M., Jun, H., Tezak, N., Schulman, J., McLeavey, C., Tworek, J., and Chen, M.
\newblock Efficient training of language models to fill in the middle.
\newblock \emph{ArXiv preprint}, abs/2207.14255, 2022.
\newblock URL \url{https://arxiv.org/abs/2207.14255}.

\bibitem[Chen et~al.(2023)Chen, Borgeaud, Irving, Lespiau, Sifre, and Jumper]{chen2023accelerating}
Chen, C., Borgeaud, S., Irving, G., Lespiau, J.-B., Sifre, L., and Jumper, J.
\newblock Accelerating large language model decoding with speculative sampling.
\newblock \emph{ArXiv preprint}, abs/2302.01318, 2023.
\newblock URL \url{https://arxiv.org/abs/2302.01318}.

\bibitem[Chen et~al.(2021)Chen, Tworek, Jun, Yuan, Pinto, Kaplan, Edwards, Burda, Joseph, Brockman, et~al.]{chen2021evaluating}
Chen, M., Tworek, J., Jun, H., Yuan, Q., Pinto, H. P. d.~O., Kaplan, J., Edwards, H., Burda, Y., Joseph, N., Brockman, G., et~al.
\newblock Evaluating large language models trained on code.
\newblock \emph{ArXiv preprint}, abs/2107.03374, 2021.
\newblock URL \url{https://arxiv.org/abs/2107.03374}.

\bibitem[Ding et~al.(2023)Ding, Wang, Ahmad, Ding, Tan, Jain, Ramanathan, Nallapati, Bhatia, Roth, and Xiang]{ding2023crosscodeeval}
Ding, Y., Wang, Z., Ahmad, W.~U., Ding, H., Tan, M., Jain, N., Ramanathan, M.~K., Nallapati, R., Bhatia, P., Roth, D., and Xiang, B.
\newblock Crosscodeeval: A diverse and multilingual benchmark for cross-file code completion.
\newblock In \emph{Thirty-seventh Conference on Neural Information Processing Systems Datasets and Benchmarks Track}, 2023.
\newblock URL \url{https://arxiv.org/abs/2310.11248}.

\bibitem[Ding et~al.(2024)Ding, Wang, Ahmad, Ramanathan, Nallapati, Bhatia, Roth, and Xiang]{ding2022cocomic}
Ding, Y., Wang, Z., Ahmad, W.~U., Ramanathan, M.~K., Nallapati, R., Bhatia, P., Roth, D., and Xiang, B.
\newblock {C}o{C}o{MIC}: Code completion by jointly modeling in-file and cross-file context.
\newblock pp.\  3433--3445, May 2024.
\newblock URL \url{https://aclanthology.org/2024.lrec-main.305}.

\bibitem[Drozdov et~al.(2022)Drozdov, Wang, Rahimi, McCallum, Zamani, and Iyyer]{drozdov-etal-2022-cant}
Drozdov, A., Wang, S., Rahimi, R., McCallum, A., Zamani, H., and Iyyer, M.
\newblock You can{'}t pick your neighbors, or can you? when and how to rely on retrieval in the k{NN}-{LM}.
\newblock In \emph{Findings of the Association for Computational Linguistics: EMNLP 2022}, pp.\  2997--3007, Abu Dhabi, United Arab Emirates, December 2022. Association for Computational Linguistics.
\newblock \doi{10.18653/v1/2022.findings-emnlp.218}.
\newblock URL \url{https://aclanthology.org/2022.findings-emnlp.218}.

\bibitem[Guo et~al.(2022)Guo, Lu, Duan, Wang, Zhou, and Yin]{guo-etal-2022-unixcoder}
Guo, D., Lu, S., Duan, N., Wang, Y., Zhou, M., and Yin, J.
\newblock {U}ni{X}coder: Unified cross-modal pre-training for code representation.
\newblock In \emph{Proceedings of the 60th Annual Meeting of the Association for Computational Linguistics (Volume 1: Long Papers)}, pp.\  7212--7225, Dublin, Ireland, May 2022. Association for Computational Linguistics.
\newblock \doi{10.18653/v1/2022.acl-long.499}.
\newblock URL \url{https://aclanthology.org/2022.acl-long.499}.

\bibitem[He et~al.(2021)He, Neubig, and Berg-Kirkpatrick]{he-etal-2021-efficient}
He, J., Neubig, G., and Berg-Kirkpatrick, T.
\newblock Efficient nearest neighbor language models.
\newblock In \emph{Proceedings of the 2021 Conference on Empirical Methods in Natural Language Processing}, pp.\  5703--5714, Online and Punta Cana, Dominican Republic, November 2021. Association for Computational Linguistics.
\newblock \doi{10.18653/v1/2021.emnlp-main.461}.
\newblock URL \url{https://aclanthology.org/2021.emnlp-main.461}.

\bibitem[Hellendoorn \& Devanbu(2017)Hellendoorn and Devanbu]{hellendoorn2017deep}
Hellendoorn, V.~J. and Devanbu, P.~T.
\newblock Are deep neural networks the best choice for modeling source code?
\newblock In Bodden, E., Sch{\"{a}}fer, W., van Deursen, A., and Zisman, A. (eds.), \emph{Proceedings of the 2017 11th Joint Meeting on Foundations of Software Engineering, {ESEC/FSE} 2017, Paderborn, Germany, September 4-8, 2017}, pp.\  763--773. {ACM}, 2017.
\newblock \doi{10.1145/3106237.3106290}.
\newblock URL \url{https://doi.org/10.1145/3106237.3106290}.

\bibitem[Hill \& Rideout(2004)Hill and Rideout]{hill2004automatic}
Hill, R. and Rideout, J.
\newblock Automatic method completion.
\newblock In \emph{19th {IEEE} International Conference on Automated Software Engineering {(ASE} 2004), 20-25 September 2004, Linz, Austria}, pp.\  228--235. {IEEE} Computer Society, 2004.
\newblock \doi{10.1109/ASE.2004.10034}.
\newblock URL \url{https://doi.ieeecomputersociety.org/10.1109/ASE.2004.10034}.

\bibitem[Jaccard(1912)]{jaccard1912distribution}
Jaccard, P.
\newblock The distribution of the flora in the alpine zone.1.
\newblock \emph{New Phytologist}, 11:\penalty0 37--50, 1912.
\newblock URL \url{https://api.semanticscholar.org/CorpusID:85574559}.

\bibitem[Jain et~al.(2023)Jain, Zhang, Ahmad, Wang, Nan, Li, Tan, Nallapati, Ray, Bhatia, Ma, and Xiang]{jain-etal-2023-contraclm}
Jain, N., Zhang, D., Ahmad, W.~U., Wang, Z., Nan, F., Li, X., Tan, M., Nallapati, R., Ray, B., Bhatia, P., Ma, X., and Xiang, B.
\newblock {C}ontra{CLM}: Contrastive learning for causal language model.
\newblock In \emph{Proceedings of the 61st Annual Meeting of the Association for Computational Linguistics (Volume 1: Long Papers)}, pp.\  6436--6459, Toronto, Canada, July 2023. Association for Computational Linguistics.
\newblock \doi{10.18653/v1/2023.acl-long.355}.
\newblock URL \url{https://aclanthology.org/2023.acl-long.355}.

\bibitem[Jiang et~al.(2023)Jiang, Xu, Gao, Sun, Liu, Dwivedi-Yu, Yang, Callan, and Neubig]{jiang2023active}
Jiang, Z., Xu, F., Gao, L., Sun, Z., Liu, Q., Dwivedi-Yu, J., Yang, Y., Callan, J., and Neubig, G.
\newblock Active retrieval augmented generation.
\newblock In Bouamor, H., Pino, J., and Bali, K. (eds.), \emph{Proceedings of the 2023 Conference on Empirical Methods in Natural Language Processing}, pp.\  7969--7992, Singapore, December 2023. Association for Computational Linguistics.
\newblock \doi{10.18653/v1/2023.emnlp-main.495}.
\newblock URL \url{https://aclanthology.org/2023.emnlp-main.495}.

\bibitem[Kadavath et~al.(2022)Kadavath, Conerly, Askell, Henighan, Drain, Perez, Schiefer, Hatfield-Dodds, DasSarma, Tran-Johnson, et~al.]{kadavath2022language}
Kadavath, S., Conerly, T., Askell, A., Henighan, T., Drain, D., Perez, E., Schiefer, N., Hatfield-Dodds, Z., DasSarma, N., Tran-Johnson, E., et~al.
\newblock Language models (mostly) know what they know.
\newblock \emph{ArXiv preprint}, abs/2207.05221, 2022.
\newblock URL \url{https://arxiv.org/abs/2207.05221}.

\bibitem[Kocetkov et~al.(2022)Kocetkov, Li, Allal, Li, Mou, Ferrandis, Jernite, Mitchell, Hughes, Wolf, et~al.]{kocetkov2022stack}
Kocetkov, D., Li, R., Allal, L.~B., Li, J., Mou, C., Ferrandis, C.~M., Jernite, Y., Mitchell, M., Hughes, S., Wolf, T., et~al.
\newblock The stack: 3 tb of permissively licensed source code.
\newblock \emph{ArXiv preprint}, abs/2211.15533, 2022.
\newblock URL \url{https://arxiv.org/abs/2211.15533}.

\bibitem[Kwon et~al.(2023)Kwon, Li, Zhuang, Sheng, Zheng, Yu, Gonzalez, Zhang, and Stoica]{kwon2023efficient}
Kwon, W., Li, Z., Zhuang, S., Sheng, Y., Zheng, L., Yu, C.~H., Gonzalez, J., Zhang, H., and Stoica, I.
\newblock Efficient memory management for large language model serving with pagedattention.
\newblock In Flinn, J., Seltzer, M.~I., Druschel, P., Kaufmann, A., and Mace, J. (eds.), \emph{Proceedings of the 29th Symposium on Operating Systems Principles, {SOSP} 2023, Koblenz, Germany, October 23-26, 2023}, pp.\  611--626. {ACM}, 2023.
\newblock \doi{10.1145/3600006.3613165}.
\newblock URL \url{https://doi.org/10.1145/3600006.3613165}.

\bibitem[Levenshtein et~al.(1966)]{levenshtein1966binary}
Levenshtein, V.~I. et~al.
\newblock Binary codes capable of correcting deletions, insertions, and reversals.
\newblock In \emph{Soviet physics doklady}, volume~10, pp.\  707--710. Soviet Union, 1966.
\newblock URL \url{https://nymity.ch/sybilhunting/pdf/Levenshtein1966a.pdf}.

\bibitem[Li et~al.(2023{\natexlab{a}})Li, Tang, Zhao, Wang, Nie, and Wen]{li-etal-2023-web}
Li, J., Tang, T., Zhao, W.~X., Wang, J., Nie, J.-Y., and Wen, J.-R.
\newblock The web can be your oyster for improving language models.
\newblock In \emph{Findings of the Association for Computational Linguistics: ACL 2023}, pp.\  728--746, Toronto, Canada, July 2023{\natexlab{a}}. Association for Computational Linguistics.
\newblock \doi{10.18653/v1/2023.findings-acl.46}.
\newblock URL \url{https://aclanthology.org/2023.findings-acl.46}.

\bibitem[Li et~al.(2023{\natexlab{b}})Li, Allal, Zi, Muennighoff, Kocetkov, Mou, Marone, Akiki, Li, Chim, Liu, Zheltonozhskii, Zhuo, Wang, Dehaene, Davaadorj, Lamy{-}Poirier, Monteiro, Shliazhko, Gontier, Meade, Zebaze, Yee, Umapathi, Zhu, Lipkin, Oblokulov, Wang, V, Stillerman, Patel, Abulkhanov, Zocca, Dey, Zhang, Moustafa{-}Fahmy, Bhattacharyya, Yu, Singh, Luccioni, Villegas, Kunakov, Zhdanov, Romero, Lee, Timor, Ding, Schlesinger, Schoelkopf, Ebert, Dao, Mishra, Gu, Robinson, Anderson, Dolan{-}Gavitt, Contractor, Reddy, Fried, Bahdanau, Jernite, Ferrandis, Hughes, Wolf, Guha, von Werra, and de~Vries]{li2023starcoder}
Li, R., Allal, L.~B., Zi, Y., Muennighoff, N., Kocetkov, D., Mou, C., Marone, M., Akiki, C., Li, J., Chim, J., Liu, Q., Zheltonozhskii, E., Zhuo, T.~Y., Wang, T., Dehaene, O., Davaadorj, M., Lamy{-}Poirier, J., Monteiro, J., Shliazhko, O., Gontier, N., Meade, N., Zebaze, A., Yee, M., Umapathi, L.~K., Zhu, J., Lipkin, B., Oblokulov, M., Wang, Z., V, R.~M., Stillerman, J., Patel, S.~S., Abulkhanov, D., Zocca, M., Dey, M., Zhang, Z., Moustafa{-}Fahmy, N., Bhattacharyya, U., Yu, W., Singh, S., Luccioni, S., Villegas, P., Kunakov, M., Zhdanov, F., Romero, M., Lee, T., Timor, N., Ding, J., Schlesinger, C., Schoelkopf, H., Ebert, J., Dao, T., Mishra, M., Gu, A., Robinson, J., Anderson, C.~J., Dolan{-}Gavitt, B., Contractor, D., Reddy, S., Fried, D., Bahdanau, D., Jernite, Y., Ferrandis, C.~M., Hughes, S., Wolf, T., Guha, A., von Werra, L., and de~Vries, H.
\newblock Starcoder: may the source be with you!, 2023{\natexlab{b}}.
\newblock URL \url{https://doi.org/10.48550/arXiv.2305.06161}.

\bibitem[Lu et~al.(2022)Lu, Duan, Han, Guo, Hwang, and Svyatkovskiy]{lu-etal-2022-reacc}
Lu, S., Duan, N., Han, H., Guo, D., Hwang, S.-w., and Svyatkovskiy, A.
\newblock {R}e{ACC}: A retrieval-augmented code completion framework.
\newblock In \emph{Proceedings of the 60th Annual Meeting of the Association for Computational Linguistics (Volume 1: Long Papers)}, pp.\  6227--6240, Dublin, Ireland, May 2022. Association for Computational Linguistics.
\newblock \doi{10.18653/v1/2022.acl-long.431}.
\newblock URL \url{https://aclanthology.org/2022.acl-long.431}.

\bibitem[Mallen et~al.(2023)Mallen, Asai, Zhong, Das, Khashabi, and Hajishirzi]{mallen-etal-2023-trust}
Mallen, A., Asai, A., Zhong, V., Das, R., Khashabi, D., and Hajishirzi, H.
\newblock When not to trust language models: Investigating effectiveness of parametric and non-parametric memories.
\newblock In \emph{Proceedings of the 61st Annual Meeting of the Association for Computational Linguistics (Volume 1: Long Papers)}, pp.\  9802--9822, Toronto, Canada, July 2023. Association for Computational Linguistics.
\newblock \doi{10.18653/v1/2023.acl-long.546}.
\newblock URL \url{https://aclanthology.org/2023.acl-long.546}.

\bibitem[Nijkamp et~al.(2023{\natexlab{a}})Nijkamp, Hayashi, Xiong, Savarese, and Zhou]{Nijkamp2023codegen2}
Nijkamp, E., Hayashi, H., Xiong, C., Savarese, S., and Zhou, Y.
\newblock Codegen2: Lessons for training llms on programming and natural languages.
\newblock \emph{ICLR}, 2023{\natexlab{a}}.
\newblock URL \url{https://arxiv.org/abs/2305.02309}.

\bibitem[Nijkamp et~al.(2023{\natexlab{b}})Nijkamp, Pang, Hayashi, Tu, Wang, Zhou, Savarese, and Xiong]{Nijkamp2022CodeGenAO}
Nijkamp, E., Pang, B., Hayashi, H., Tu, L., Wang, H., Zhou, Y., Savarese, S., and Xiong, C.
\newblock Codegen: An open large language model for code with multi-turn program synthesis.
\newblock In \emph{The Eleventh International Conference on Learning Representations, {ICLR} 2023, Kigali, Rwanda, May 1-5, 2023}. OpenReview.net, 2023{\natexlab{b}}.
\newblock URL \url{https://openreview.net/pdf?id=iaYcJKpY2B\_}.

\bibitem[OpenAI(2023)]{OpenAI2023GPT4TR}
OpenAI.
\newblock Gpt-4 technical report.
\newblock \emph{ArXiv preprint}, abs/2303.08774, 2023.
\newblock URL \url{https://arxiv.org/abs/2303.08774}.

\bibitem[Parnas(1972)]{parnas1972criteria}
Parnas, D.~L.
\newblock On the criteria to be used in decomposing systems into modules.
\newblock \emph{Commun. {ACM}}, 15\penalty0 (12):\penalty0 1053--1058, 1972.
\newblock \doi{10.1145/361598.361623}.
\newblock URL \url{https://doi.org/10.1145/361598.361623}.

\bibitem[Pei et~al.(2023)Pei, Zhao, Lausen, Zha, and Karypis]{Pei_Zhao_Lausen_Zha_Karypis_2023}
Pei, H., Zhao, J., Lausen, L., Zha, S., and Karypis, G.
\newblock Better context makes better code language models: A case study on function call argument completion.
\newblock \emph{Proceedings of the AAAI Conference on Artificial Intelligence}, 37\penalty0 (4):\penalty0 5230--5238, Jun. 2023.
\newblock \doi{10.1609/aaai.v37i4.25653}.
\newblock URL \url{https://ojs.aaai.org/index.php/AAAI/article/view/25653}.

\bibitem[Ram et~al.(2023)Ram, Levine, Dalmedigos, Muhlgay, Shashua, Leyton-Brown, and Shoham]{ram2023ralm}
Ram, O., Levine, Y., Dalmedigos, I., Muhlgay, D., Shashua, A., Leyton-Brown, K., and Shoham, Y.
\newblock In-context retrieval-augmented language models.
\newblock \emph{Transactions of the Association for Computational Linguistics}, 11:\penalty0 1316--1331, 2023.
\newblock \doi{10.1162/tacl_a_00605}.
\newblock URL \url{https://aclanthology.org/2023.tacl-1.75}.

\bibitem[Ren et~al.(2020)Ren, Guo, Lu, Zhou, Liu, Tang, Sundaresan, Zhou, Blanco, and Ma]{ren2020codebleu}
Ren, S., Guo, D., Lu, S., Zhou, L., Liu, S., Tang, D., Sundaresan, N., Zhou, M., Blanco, A., and Ma, S.
\newblock Codebleu: a method for automatic evaluation of code synthesis.
\newblock \emph{ArXiv preprint}, abs/2009.10297, 2020.
\newblock URL \url{https://arxiv.org/abs/2009.10297}.

\bibitem[Roziere et~al.(2023)Roziere, Gehring, Gloeckle, Sootla, Gat, Tan, Adi, Liu, Remez, Rapin, et~al.]{roziere2023code}
Roziere, B., Gehring, J., Gloeckle, F., Sootla, S., Gat, I., Tan, X.~E., Adi, Y., Liu, J., Remez, T., Rapin, J., et~al.
\newblock Code llama: Open foundation models for code.
\newblock \emph{ArXiv preprint}, abs/2308.12950, 2023.
\newblock URL \url{https://arxiv.org/abs/2308.12950}.

\bibitem[Shi et~al.(2023)Shi, Min, Yasunaga, Seo, James, Lewis, Zettlemoyer, and Yih]{shi2023replug}
Shi, W., Min, S., Yasunaga, M., Seo, M., James, R., Lewis, M., Zettlemoyer, L., and Yih, W.-t.
\newblock Replug: Retrieval-augmented black-box language models.
\newblock \emph{ArXiv preprint}, abs/2301.12652, 2023.
\newblock URL \url{https://arxiv.org/abs/2301.12652}.

\bibitem[Shrivastava et~al.(2023{\natexlab{a}})Shrivastava, Kocetkov, de~Vries, Bahdanau, and Scholak]{shrivastava2023repofusion}
Shrivastava, D., Kocetkov, D., de~Vries, H., Bahdanau, D., and Scholak, T.
\newblock Repofusion: Training code models to understand your repository.
\newblock \emph{ArXiv preprint}, abs/2306.10998, 2023{\natexlab{a}}.
\newblock URL \url{https://arxiv.org/abs/2306.10998}.

\bibitem[Shrivastava et~al.(2023{\natexlab{b}})Shrivastava, Larochelle, and Tarlow]{shrivastava2023repository}
Shrivastava, D., Larochelle, H., and Tarlow, D.
\newblock Repository-level prompt generation for large language models of code.
\newblock In Krause, A., Brunskill, E., Cho, K., Engelhardt, B., Sabato, S., and Scarlett, J. (eds.), \emph{International Conference on Machine Learning, {ICML} 2023, 23-29 July 2023, Honolulu, Hawaii, {USA}}, volume 202 of \emph{Proceedings of Machine Learning Research}, pp.\  31693--31715. {PMLR}, 2023{\natexlab{b}}.
\newblock URL \url{https://proceedings.mlr.press/v202/shrivastava23a.html}.

\bibitem[Svyatkovskiy et~al.(2020)Svyatkovskiy, Deng, Fu, and Sundaresan]{svyatkovskiy2020intellicode}
Svyatkovskiy, A., Deng, S.~K., Fu, S., and Sundaresan, N.
\newblock Intellicode compose: code generation using transformer.
\newblock In Devanbu, P., Cohen, M.~B., and Zimmermann, T. (eds.), \emph{{ESEC/FSE} '20: 28th {ACM} Joint European Software Engineering Conference and Symposium on the Foundations of Software Engineering, Virtual Event, USA, November 8-13, 2020}, pp.\  1433--1443. {ACM}, 2020.
\newblock \doi{10.1145/3368089.3417058}.
\newblock URL \url{https://doi.org/10.1145/3368089.3417058}.

\bibitem[Tu et~al.(2014)Tu, Su, and Devanbu]{tu2014localness}
Tu, Z., Su, Z., and Devanbu, P.~T.
\newblock On the localness of software.
\newblock In Cheung, S., Orso, A., and Storey, M.~D. (eds.), \emph{Proceedings of the 22nd {ACM} {SIGSOFT} International Symposium on Foundations of Software Engineering, (FSE-22), Hong Kong, China, November 16 - 22, 2014}, pp.\  269--280. {ACM}, 2014.
\newblock \doi{10.1145/2635868.2635875}.
\newblock URL \url{https://doi.org/10.1145/2635868.2635875}.

\bibitem[Wang et~al.(2021)Wang, Shi, Du, Yang, Hu, Han, Zhang, and Zhang]{wang2021cocosum}
Wang, Y., Shi, E., Du, L., Yang, X., Hu, Y., Han, S., Zhang, H., and Zhang, D.
\newblock Cocosum: Contextual code summarization with multi-relational graph neural network.
\newblock \emph{ArXiv preprint}, abs/2107.01933, 2021.
\newblock URL \url{https://arxiv.org/abs/2107.01933}.

\bibitem[Wang et~al.(2023)Wang, Li, Sun, and Liu]{wang2023self}
Wang, Y., Li, P., Sun, M., and Liu, Y.
\newblock Self-knowledge guided retrieval augmentation for large language models.
\newblock In Bouamor, H., Pino, J., and Bali, K. (eds.), \emph{Findings of the Association for Computational Linguistics: EMNLP 2023}, pp.\  10303--10315, Singapore, 2023. Association for Computational Linguistics.
\newblock \doi{10.18653/v1/2023.findings-emnlp.691}.
\newblock URL \url{https://aclanthology.org/2023.findings-emnlp.691}.

\bibitem[Ye \& Fischer(2002)Ye and Fischer]{ye2002supporting}
Ye, Y. and Fischer, G.
\newblock Supporting reuse by delivering task-relevant and personalized information.
\newblock In Tracz, W., Young, M., and Magee, J. (eds.), \emph{Proceedings of the 24th International Conference on Software Engineering, {ICSE} 2002, 19-25 May 2002, Orlando, Florida, {USA}}, pp.\  513--523. {ACM}, 2002.
\newblock \doi{10.1145/581339.581402}.
\newblock URL \url{https://doi.org/10.1145/581339.581402}.

\bibitem[Zan et~al.(2022)Zan, Chen, Lin, Guan, Yongji, and Lou]{zan-etal-2022-language}
Zan, D., Chen, B., Lin, Z., Guan, B., Yongji, W., and Lou, J.-G.
\newblock When language model meets private library.
\newblock In \emph{Findings of the Association for Computational Linguistics: EMNLP 2022}, pp.\  277--288, Abu Dhabi, United Arab Emirates, December 2022. Association for Computational Linguistics.
\newblock \doi{10.18653/v1/2022.findings-emnlp.21}.
\newblock URL \url{https://aclanthology.org/2022.findings-emnlp.21}.

\bibitem[Zhang et~al.(2023)Zhang, Chen, Zhang, Keung, Liu, Zan, Mao, Lou, and Chen]{zhang2023repocoder}
Zhang, F., Chen, B., Zhang, Y., Keung, J., Liu, J., Zan, D., Mao, Y., Lou, J.-G., and Chen, W.
\newblock {R}epo{C}oder: Repository-level code completion through iterative retrieval and generation.
\newblock In Bouamor, H., Pino, J., and Bali, K. (eds.), \emph{Proceedings of the 2023 Conference on Empirical Methods in Natural Language Processing}, pp.\  2471--2484, Singapore, 2023. Association for Computational Linguistics.
\newblock \doi{10.18653/v1/2023.emnlp-main.151}.
\newblock URL \url{https://aclanthology.org/2023.emnlp-main.151}.

\bibitem[Zhou et~al.(2023)Zhou, Alon, Xu, Jiang, and Neubig]{zhou23docprompting}
Zhou, S., Alon, U., Xu, F.~F., Jiang, Z., and Neubig, G.
\newblock Docprompting: Generating code by retrieving the docs.
\newblock In \emph{The Eleventh International Conference on Learning Representations, {ICLR} 2023, Kigali, Rwanda, May 1-5, 2023}. OpenReview.net, 2023.
\newblock URL \url{https://openreview.net/pdf?id=ZTCxT2t2Ru}.

\end{thebibliography}
